\documentstyle[aaspp4,tighten,flushrt]{article}

\newcommand{\beq}{\begin{equation}}
\newcommand{\eeq}{\end{equation}}
\newcommand{\beqa}{\begin{eqnarray}}
\newcommand{\eeqa}{\end{eqnarray}}

\newcommand{\lexp}{\mathop{\langle}}
\newcommand{\rexp}{\mathop{\rangle}}
\newcommand{\rexpc}{\mathop{\rangle_c}}

\def\d{\delta}

\def\dD{\delta_{\rm D}}

\font\BF=cmmib10
\def\k{{\hbox{\BF k}}}

\def\x{{\hbox{\BF x}}}
\def\y{{\hbox{\BF y}}}
\def\tk{\hat k}

\def\la{\mathrel{\mathpalette\fun <}}
\def\ga{\mathrel{\mathpalette\fun >}}
\def\fun#1#2{\lower3.6pt\vbox{\baselineskip0pt\lineskip.9pt
        \ialign{$\mathsurround=0pt#1\hfill##\hfil$\crcr#2\crcr\sim\crcr}}}

\begin{document}

\title{How Many Galaxies Fit in a Halo? \\ Constraints on
Galaxy Formation Efficiency from Spatial Clustering}

\author{Rom\'an Scoccimarro\altaffilmark{1}, Ravi K. 
Sheth\altaffilmark{2}, Lam Hui\altaffilmark{1}, and Bhuvnesh
Jain\altaffilmark{3}}

\altaffiltext1{Institute for Advanced Study, School of Natural Sciences,
Einstein Drive, Princeton, NJ 08540}

\altaffiltext2{NASA/Fermilab Astrophysics Center, Fermi National 
Accelerator Laboratory, Batavia, IL  60510}

\altaffiltext3{Department of Physics, Johns Hopkins University,
Baltimore, MD 21218}

\begin{abstract}

We study galaxy clustering in the framework of halo models, where
gravitational clustering is described in terms of dark matter
halos. At small scales, dark matter clustering statistics are
dominated by halo density profiles, whereas at large scales,
correlations are the result of combining non-linear perturbation
theory with halo biasing. Galaxies are assumed to follow the dark
matter profiles of the halo they inhabit, and galaxy formation
efficiency is characterized by the number of galaxies that populate a
halo of given mass. This approach leads to generic predictions: the
galaxy power spectrum shows a power-law behavior even though the dark
matter does not, and the galaxy higher-order correlations show smaller
amplitudes at small scales than their dark matter counterparts.  Both
are in qualitatively agreement with measurements in galaxy
catalogs. We find that requiring the model to fit both the second and
third order moments of the APM galaxies provides a strong constraint
on galaxy formation models.  The data at large scales require that
galaxy formation be relatively efficient at small masses, $m \approx
10^{10} M_{\sun}/h$, whereas data at smaller scales require that the
number of galaxies in a halo scale approximately as the mass to the
$0.8$th power in the high-mass limit. These constraints are independent
of those derived from the luminosity function or Tully-Fisher
relation. We also predict the power spectrum, bispectrum, and
higher-order moments of the mass density field in this framework.
Although halo models agree well with measurements of the mass power
spectrum and the higher order $S_p$ parameters in N-body simulations,
the model assumption that halos are spherical leads to disagreement in
the configuration dependence of the bispectrum at small scales.  We
stress the importance of finite volume effects in higher-order
statistics and show how they can be estimated in this approach.

\end{abstract}

\clearpage

\section{Introduction}

Understanding galaxy clustering is one of the main goals of cosmology.
The wealth of information provided by galaxy surveys can only be used
to extract useful cosmological information if we understand the
relation between galaxy and dark matter clustering---biasing.  On
large enough scales, galaxy biasing can be described as a local
process, and so galaxy clustering can be used as a direct probe of the
primordial spectrum and Gaussianity of initial conditions (Fry \&
Gazta\~naga 1993; Frieman \& Gazta\~naga 1994; Fry 1994; Fry \&
Scherrer 1994; Gazta\~naga \& M\"ah\"onen 1996; Matarrese, Verde \&
Heavens 1997; Gazta\~naga \& Fosalba 1998; Frieman \& Gazta\~naga
1999; Scoccimarro et al. 2000; Durrer et al. 2000). On smaller scales,
however, non-negligible contributions from complex astrophysical
processes relevant to galaxy formation may complicate the description
of galaxy biasing.

Galaxy formation is not yet understood from first physical principles.
However, following White \& Rees (1978) and White \& Frenk (1991), a
number of prescriptions based on reasonable recipes for approximating
the complicated physics have been proposed for incorporating galaxy
formation into numerical simulations of dark matter gravitational
clustering (see, e.g., Kauffmann et al. 1999, Somerville \& Primack
1999 or Benson et al. 2000 for some of the most recent work).  These
``semianalytic galaxy formation'' schemes can provide detailed
predictions for galaxy properties in hierarchical structure formation
models, which can then be compared with observations.

The basic assumption in the semianalytic approach is that galaxy
biasing can be described as a two-step process. First, the formation
and clustering of dark matter halos can be modeled by neglecting
non-gravitational effects.\footnote{Halos here are defined in the
sense of Press \& Schechter (1974).  There is not necessarily a
one-to-one correspondence between halos and galaxies.}  This can be
done reasonably accurately following the analytic results of Mo \&
White (1996), Mo, Jing \& White (1997) and Sheth \& Lemson (1999).
Second, the distribution of galaxies within halos, which in principle
depends on complicated physics, can be described by a number of
simplifying assumptions regarding gas cooling and feedback effects
from supernova. For the purposes of this paper, the main outcome of
this second step is the number of galaxies that populate a halo of a
given mass, $N_{\rm gal}(m)$.

In this paper we consider the problem of galaxy clustering from a
complementary point of view to semianalytic models. We construct
clustering statistics from properties of dark matter halos and the
$N_{\rm gal}(m)$ relation, and show how these simple ingredients can
be put together to make reasonably accurate analytic predictions about
the galaxy power spectrum, bispectrum, and higher-order moments of the
galaxy field. We also consider the inverse problem:  we show how 
measurements of galaxy clustering can constrain the $N_{\rm gal}(m)$ 
relation.  We show in particular that the variance and skewness of 
the galaxy distribution in the APM survey provide significant 
constraints on the $N_{\rm gal}(m)$ relation.

Our approach to gravitational clustering has a long history, dating
back to Neyman, \& Scott (1952), and then explored further by Peebles
(1974), and McClelland \& Silk (1977a,b;1978). These works considered
perturbations described by halos of a given size and profile, but
distributed at random. A complete treatment which includes the effects
of halo-halo correlations was first given by Scherrer \& Bertschinger
(1991). Recent work has focused on applications of this formalism to
the clustering of dark matter, e.g. the small-scale behavior of the
two-point correlation function (Sheth \& Jain 1997), the power
spectrum (Seljak 2000; Peacock \& Smith 2000; Ma \& Fry 2000; Cooray
\& Hu 2000), and the bispectrum for equilateral configurations (Ma \&
Fry 2000; Cooray \& Hu 2000).  Interest in this approach has been
undoubtedly sparked by recent results from numerical simulations on
the properties of dark matter halos (Navarro, Frenk, \& White 1996,
1997; Bullock et al. 1999; Moore et al. 1999).

This paper is organized as follows. In Section 2 we review the halo
model formalism for the power spectrum, bispectrum and higher-order
moments of the smoothed density field, and compare the predictions
with numerical simulations in Section 3. We discuss in detail the role
of finite volume effects which, if neglected, can lead to incorrect
conclusions regarding higher-order statistics.  In Section 4, we use
the $N_{\rm gal}(m)$ relation to make predictions for galaxy, rather
than dark matter, clustering statistics (also see Jing, Mo \& B\"orner
1998; Seljak 2000; Peacock \& Smith 2000).  To illustrate our
approach, we use the $N_{\rm gal}(m)$ relation obtained from the
semianalytic models of Kauffmann et al. (1999).  In Section 5 we
discuss the constraints on $N_{\rm gal}(m)$ derived from analysis of
counts-in-cells of the APM survey.  Section 6 summarizes our
conclusions.

\section{Dark Matter Clustering} 

\subsection{Formalism} 

In this section we follow the formalism developed by Scherrer \&
Bertschinger (1991). The dark matter density field is written as

\beq
\rho(\x) = \sum_{i} f(\x-\x_i,m_i) \equiv \sum_{i} m_i\ u(\x-\x_i,m_i)
= \sum_i \int dm d^3 x' \delta (m-m_i) \delta^3 (\x'-\x_i) m u(\x-\x',m),
\eeq

\noindent where $f$ denotes the density profile of a halo of mass $m_i$
located at position $\x_i$. The mean density is 

\beq 
\bar{\rho} = \lexp \rho(\x) \rexp =\lexp \sum_{i} m_i\ u(\x-\x_i,m_i)
\rexp = \int n(m) m dm \int d^3x' u_m(\x-\x'),
\eeq

\noindent where we have replaced the ensemble average by the average
over the mass function $n(m)$ (which gives the density of halos per
unit mass) and an average over space i.e. $\langle \sum_i \delta
(m-m_i) \delta^3 (\x'-\x_i)\rangle = n(m)$.  Our normalization
convention is such that $\int d^3x' u_m(\x-\x')=1$ and $\int n(m) m
dm=\bar{\rho}$. The two-point correlation function can be written as

\beqa
\bar{\rho}^2 \xi(\x-\x')&=& 
\int n(m) m^2 dm \int d^3y u_m(\y) u_m(\y+\x-\x')
+ \int n(m_1) m_1 dm_1 \int n(m_2) m_2 dm_2 \nonumber \\ && \times 
\int d^3x_1 u_{m_1}(\x-\x_1) \int d^3x_2 u_{m_2}(\x'-\x_2) 
\ \xi(\x_1-\x_2;m_1,m_2),
\eeqa

\noindent where the first term describes the case where the two
particles occupy the same halo, and the second term represents the
contribution of particles in different halos, with
$\xi(\x-\x';m_1,m_2)$ being the two-point correlation function of
halos of mass $m_1$ and $m_2$. Since we are dealing with convolutions
of halo profiles, it is much easier to work in Fourier space, where
expressions become multiplications over the Fourier transform of halo
profiles. We use the following Fourier space conventions:

\beq
A(\k) = \int \frac{d^3 x}{(2\pi)^3} \exp(-i \k \cdot \x) A(\x),
\eeq
and 
\beq
\lexp \d(\k_1) \d(\k_2) \rexp = \dD(\k_{12}) P(k_1), 
\eeq
\beq
\lexp \d(\k_1) \d(\k_2) \d(\k_3) \rexp = \dD(\k_{123}) B_{123}, 
\eeq

\noindent where $P(k)$ and $B_{123}\equiv B(k_1,k_2,k_3)$ denote the
power spectrum and bispectrum, respectively. Thus, the power spectrum
reads

\beq 
\bar{\rho}^2 P(k)= (2\pi)^3 \int n(m) m^2 dm |u_m(\k)|^2 +
(2\pi)^6 \int u_{m_1}(k) n(m_1) m_1 dm_1 \int u_{m_2}(k) 
n(m_2) m_2 dm_2 P(k;m_1,m_2),
\label{pkm}
\eeq

\noindent where $P(k;m_1,m_2)$ represents the power spectrum of halos
of mass $m_1$ and $m_2$. Similarly, the bispectrum is given by 

\beqa 
\bar{\rho}^3 B_{123}&=& (2\pi)^3 \int n(m) m^3 dm\
 \Pi_{i=1}^3 u_m(\k_i) +  (2\pi)^6 \int u_{m_1}(k_1)
n(m_1) m_1 dm_1 \int u_{m_2}(k_2) u_{m_2}(k_3) n(m_2) m_2^2 dm_2
\nonumber \\ 
& & \times P(k_1;m_1,m_2) + {\rm cyc.} 
+ (2\pi)^9 \Big( \prod_{i=1}^3 \int u_{m_i}(k_i) n(m_i) m_i dm_i \Big)
  B_{123}(m_1,m_2,m_3) , 
\eeqa

\noindent where $B_{123}(m_1,m_2,m_3)$ denotes the bispectrum of halos
of mass $m_1,m_2,m_3$. So far the treatment has been completely
general. To make the model quantitative, we must specify 
the halo profile $u_m(\x)$, the halo mass function $n(m)$ and 
the halo-halo correlations encoded in $P(k;m_1,m_2)$,
$B_{123}(m_1,m_2,m_3)$, etc.

\subsection{Halo Profiles} 

For the halo profile we use (Navarro, Frenk \& White 1997;
hereafter NFW)

\beq
u_R(r) = \frac{f c^3}{4\pi R_{vir}^3} \frac{1}{cr/R_{vir} 
(1+cr/R_{vir})^2},
\eeq

\noindent where $f=1/[\ln(1+c)-c/(1+c)]$, $R_{vir}$ is the virial
radius of the halo, related to its mass by $m=(4\pi R_{vir}^3/3)\Delta
\bar{\rho}$, where $\Delta=200,340$ for an $\Omega=1,0.3$ universe,
respectively. We will also use the Lagrangian radius $R$ (the initial
radius where the mass $m$ came from) to specify halo sizes; $R=R_{vir}
\Delta^{1/3}$. It is convenient to work in units of the characteristic
non-linear mass $m_*$ or the equivalent scale $R_*$ (defined such that
$\sigma(R_*)=\d_c$; note that $m_*=(4\pi R_*^3/3)\bar{\rho}$). Since
$m=1.16\times 10^{12} \Omega (R {\rm h/Mpc})^3 M_{\sun}$/h, for
$\Lambda$CDM ($\Omega=0.3$, $\Omega_\Lambda=0.7$) with
$\sigma_8=0.90$, $R_*=3.135$ Mpc/h, so $m_*=1.07\times 10^{13}
M_{\sun}$/h. The Fourier transform of the halo profile reads:

\beq
u_R(\k) = \int \frac{d^3x}{(2\pi)^3} \exp(-i \k \cdot \x)
u_R(r) \equiv \frac{1}{(2\pi)^3} u(\tk,y),
\label{uft}
\eeq

\noindent where $y=R/R_*$, $\tk=kR_*\Delta^{-1/3}$, and

\beq
u(\tk,y)= f \ 
\Big[ \sin \kappa \Big( {\rm Si}[\kappa(1+c)]- {\rm Si}(\kappa) \Big)
+ \cos \kappa \Big( {\rm Ci}[\kappa(1+c)]- {\rm Ci}(\kappa) \Big) -
\frac{\sin (\kappa c)}{\kappa (1+c)} \Big], 
\label{uNFW}
\eeq

\noindent where $\kappa \equiv \tk y/c$, ${\rm Si}(x) = \int_0^x dt \
\sin (t)/t$ is the sine integral and ${\rm Ci}(x) = - \int_x^\infty dt
\ \cos (t)/t$ is the cosine integral function. The concentration
parameter quantifies the transition from the inner to the outer slope
in the NFW profile.  For the concentration parameter of halos, we use
the result (Bullock et al. 1999)

\beq
c(m) \approx 9 \Big( \frac{m}{m_*} \Big)^{-0.13}.
\label{conc}
\eeq

Note that halos defined as above are somewhat different from the
prescription in NFW, which defines $R_{vir}$ as the scale within which
the mean enclosed density is 200 times the critical density,
independent of cosmology. Both approaches agree for $\Omega=1$,
whereas for the model we use $\Omega=0.3$ and thus our virial radius
is defined at the scale where the mean enclosed density density is
$340\times 0.3 \approx 100$ times the critical density. Thus, our
characteristic density is smaller than NFW's by a factor of two, and
our virial radius is correspondingly larger. These two effects
partially cancel and lead to a very similar prediction for halo
profiles (taking also into account that the concentration parameters
are slightly different as well).

\subsection{Mass Function} 

The mass function in normalized units reads

\beq n(m)\ m\ dm \equiv \bar{\rho}\ \frac{dy}{y}\ \tilde n(y)=
\bar{\rho}\ \frac{dy}{y} \ \gamma\,A{\sqrt{\alpha\nu^3\over
2\pi}}\Big(1+(\alpha \nu^2)^{-p}\Big) \exp(-\alpha \nu^2/2) , \eeq

\noindent where $\nu \equiv \d_c/\sigma$ with $\d_c=1.68$ the colapse
threshold given by the spherical collapse model, $A=0.5,0.322$,
$p=0,0.3$ and $\alpha=1,0.707$ for the PS (Press \& Schechter 1974)
and ST (Sheth \& Tormen 1999) mass function, respectively (see Jenkins
et al. 2000 for a recent comparison of these mass functions against
N-body simulations). Note that in this formula the linear variance is
$\sigma^2= \sigma^2_L(R_* y)$, and $\gamma(R)\equiv
-d\ln\sigma_L^2(R)/d\ln R$ is the logarithmic slope of the linear
variance as a function of scale.

\subsection{Halo-Halo Correlations} 

Following Mo \& White (1996) (see also Mo, Jing \& White (1997); Sheth
\& Lemson (1999); Sheth \& Tormen (1999) for extensions), halo-halo
correlations are described by non-linear perturbation theory plus a
halo biasing prescription obtained from the spherical collapse
model\footnote{A treatment of halo biasing beyond the spherical
collapse approximation using perturbation theory is given in Catelan
et al. (1998)}. For the PS and ST mass functions, we will need

\beq
b_1(m) = 1 + \epsilon_1 + E_1,
\label{b1}
\eeq
\beq
b_2(m) = 2(1+a_2) (\epsilon_1 + E_1) + \epsilon_2 + E_2,
\label{b2}
\eeq 
\beq
b_3(m) = 6(a_2 + a_3)\,(\epsilon_1+E_1) 
         + 3(1 + 2a_2)\,(\epsilon_2 + E_2) + \epsilon_3 + E_3,
\label{b3}
\eeq 
\beq
b_4(m) = 24(a_3 + a_4)(\epsilon_1+E_1) 
         + 12\Big[a_2^2 + 2(a_2 + a_3)\Big](\epsilon_2+E_2)
         + 4(1 + 3a_2)(\epsilon_3+E_3) + \epsilon_4 + E_4,
\label{b4}
\eeq 

\noindent where 
\beqa
\epsilon_1&=& \frac{\alpha \nu^2-1}{\d_c},\ \ \ \ 
\epsilon_2 = {\alpha\nu^2\over\d_c^2} (\alpha \nu^2-3),\ \ \ \ 
\epsilon_3 = {\alpha\nu^2\over\d_c^3}(\alpha^2\nu^4 - 6\alpha\nu^2 + 3),\nonumber\\
\epsilon_4&=& \left({\alpha\nu^2\over\d_c^2}\right)^2
              (\alpha^2\nu^4 - 10\alpha\nu^2 + 15) , \nonumber \\
E_1 &=& \frac{2p/\d_c}{1+ (\alpha \nu^2)^p},\ \ \ \ 
{E_2\over E_1} = \left({1+2p\over\d_c} + 2\epsilon_1\right) \ \ \ \
{E_3\over E_1} = \left({4(p^2-1) + 6p\alpha\nu^2\over\d_c^2} + 3\epsilon_1^2\right),\nonumber \\
{E_4\over E_1} &=& {2\alpha\nu^2\over\d_c^2} 
\left(2{\alpha^2\nu^4\over\d_c} - 15\epsilon_1\right)
 + 2{(1 + p)\over\d_c^2} 
 \left({4(p^2-1) + 8(p-1)\alpha\nu^2 + 3\over\d_c} +
6\alpha\nu^2\epsilon_1\right)\nonumber \\ \nonumber \\
a_2 &=& -17/21,\ \ \ \ 
a_3 = 341/567,\ \ \ \ {\rm and}\ \ \ \ 
a_4 = -55805/130977.
\eeqa

\noindent All the $E_n$'s are zero, and $\alpha=1$, if $n(m)$ is given 
by the PS formula.  In this case, our formulae reduce to those in 
Mo, Jing \& White (1997), although our expression for $b_4(m)$ 
corrects a typographical error in their equation~(15c).

\noindent Halo-halo correlations read

\beq
P(k;m_1,m_2)=b_1(m_1) b_1(m_2) P_L(k),
\eeq
\beq
B_{123}(m_1,m_2,m_3)= b_1(m_1) b_1(m_2) b_1(m_3) B_{123}^{\rm PT} + 
b_1(m_1) b_1(m_2)  b_2(m_3) P_L(k_1) P_L(k_2) + {\rm cyc.}, 
\eeq

\noindent and similarly for higher-order moments [see
Eqs.(\ref{S3S4g}-\ref{S5g2})]. The symbol $P_L(k)$ and $B_{123}^{\rm
PT}$ denotes respectively the linear power spectrum and the
second-order perturbative bispectrum (Fry 1984)

\beq
B_{123}^{\rm PT}= 2F_2(\k_1,\k_2) P_L(k_1) P_L(k_2) + {\rm cyc.}, 
\eeq

\noindent where $F_2(\k_1,\k_2)=5/7+1/2 \cos \theta_{12}
(k_1/k_2+k_2/k_1) + 2/7 \cos^2 \theta_{12}$, with $\k_1 \cdot \k_2=
k_1 k_2 \cos \theta_{12}$.  By construction, the bias parameters obey
($n=2,3, \ldots$)

\beq \int \frac{dy}{y} \tilde n(y) b_1(y)=1,\ \ \ \ \ \int
\frac{dy}{y} \tilde n(y) b_n(y)=0.
\label{biascon}
\eeq

\subsection{Results} 

Using the ingredients above, we can rewrite the power spectrum and
bispectrum as 

\beq
P(k)= \Big[ M_{11}(\tk)\Big ]^2  P_L(k) + 
M_{02}(\tk,\tk) ,
\label{power}
\eeq

\beqa
B_{123}&=& \Big( \prod_{i=1}^3 M_{11}(\tk_i) \Big) 
B_{123}^{\rm PT} + \Big(M_{11}(\tk_1)M_{11}(\tk_2)M_{21}(\tk_3) 
P_L(k_1) P_L(k_2) +  {\rm cyc.} \Big) \nonumber
\\
& & + \Big( M_{11}(\tk_1)M_{12}(\tk_2,\tk_3) 
P_L(k_1) + {\rm cyc.} \Big) 
+  M_{03}(\tk_1,\tk_2,\tk_3)
\label{bisp}
\eeqa

\noindent where ($b_0\equiv 1$)

\beq
M_{ij}(\tk_1,\ldots,\tk_j) \equiv 
\int \frac{dy}{y} \tilde n(y) b_i(y) [u(\tk_1,y)\ldots u(\tk_j,y)]
\Big(\frac{R_*^3 y^3}{6\pi^2} \Big)^{j-1}.
\label{Iij}
\eeq

\noindent It is convenient to define the reduced bispectrum $Q_{123}$,

\beq
Q_{123} \equiv \frac{B_{123}}{P_1P_2+P_2P_3+P_3P_1},
\label{Q}
\eeq

\noindent which shows a much weaker scale dependence than the
bispectrum itself, since at large scales PT predicts $B \propto P^2$,
and at small scales the hierarchical ansatz also predicts such a
behavior. 

We can also obtain the one-point moments smoothed at scale $R$ from 
Fourier space correlations by integrating with a top-hat window function 
in Fourier space, $W(kR)$. For example, from Eq.~(\ref{power}), the 
variance is

\beq
\sigma^2(R)= \int d^3k P(k) W(kR)^2 \approx 
\sigma^2_L (R) + \int \frac{dy}{y} \tilde n(y) y^3\ \overline{u^2}(R,y),
\eeq

\noindent where 
\beq
\overline{u^m}(R,y) = \frac{2 \Delta}{3\pi} \int
\tk^2 d\tk\ [u(\tk,y)]^m\ W^2(k R),
\eeq

\noindent and we have assumed that $\int M_{11}^2(\tk)P_L(k) d^3k
W(kR)^2 \approx \sigma^2_L (R)$ since $M_{11}(\tk) \rightarrow 1$ as
$k \rightarrow 0$, and at smaller scales the power spectrum is
dominated by the second term in Eq.~(\ref{power}). Similarly the third
moment reads ($W_i\equiv W(k_iR)$),

\beqa
\lexp \d^3(R) \rexp &=& \int d^3k_1 d^3k_2 d^3k_3 \d_D(\k_{123}) B_{123}
W_1 W_2 W_3 \nonumber \\
& \approx  & S_3^{\rm PT} \sigma^4_L (R) + 3 \sigma^2_L (R)
\int \frac{dy}{y} \tilde n(y) b_1(y) y^3\ \overline{u^2}(R,y) + 
\int \frac{dy}{y} \tilde n(y) y^6 \overline{u}(R,y) \overline{u^2}(R,y).
\label{d3deriv}
\eeqa

\noindent There are several approximations involved in this result. 
First, we take the large-scale limit $M_{11} \approx 1$,
$M_{21} \approx 0$, valid to a good approximation because of the
consistency conditions, Eq.~(\ref{biascon}). In addition, we assume
that the configuration dependence of the 1-halo and 2-halo terms in
Eq.(\ref{bisp}) can be neglected (this holds very well for the 2-halo
term and approximately for the 1-halo term, as we shall discuss below;
e.g. see bottom right panel in Fig.~\ref{fig_Qnb_halo_config}). We can
thus evaluate these terms for equilateral configurations, and simplify
the angular integration by further assuming $W_1 W_2 W_{12} \approx
W_1^2 W_2^2$, the leading-order term in the multipolar expansion. With
similar approximations, we can derive higher-order connected
moments. Define

\beq
A_{ij}(R) \equiv \int \frac{dy}{y} \tilde n(y)\ b_i(y)\ y^{3(j+1)}\ 
[\overline{u}(R,y)]^j\  \overline{u^2}(R,y),
\label{Aij}
\eeq

\noindent so that $\sigma^2 = \sigma_L^2 +A_{00}$ and $\lexp \d^3
\rexp = S_3^{\rm PT} \sigma^4_L + 3 \sigma^2_L A_{10} +A_{01}$. Then
it follows that

\beq
\lexp \d^4 \rexpc  =
S_4^{\rm PT} \sigma^6_L + 6 \frac{S_3^{\rm PT}}{3} \sigma^4_L A_{10} + 7
\frac{4\sigma^2_L}{7} A_{11} +A_{02}, 
\label{S4} 
\eeq

\beq
\lexp \d^5 \rexpc =
S_5^{\rm PT} \sigma^8_L + 10 \frac{S_4^{\rm PT}}{16} \sigma^6_L A_{10} + 
25 \frac{3S_3^{\rm PT}}{5} \sigma^4_L A_{11} + 15
\frac{\sigma^2_L}{3}  A_{12} +A_{03},
\label{S5} 
\eeq

\noindent where the terms in $\lexp \d^n \rexpc$ are ordered from
$n$-halo to 1-halo contributions. The coefficient of an $m$-halo
contribution to $\lexp \d^n \rexpc$ is given by $s(n,m)$ (e.g. $6$ and
$7$ in the second and third terms of Eq. [\ref{S4}]), the Stirling
number of second kind, which is the number of ways of putting $n$
distinguishable objects ($\d$) into $m$ cells (halos), with no cells
empty (Scherrer \& Bertschinger 1991). Thus, in general we can write

\beq \lexp \d^n \rexpc = S_n^{\rm PT} \sigma^{2(n-1)}_L +
\sum_{m=2}^{n-1} s(n,m)\ \alpha_{nm} S_{m}^{\rm PT}\ \sigma^{2(m-1)}_L
A_{1n-m-1} + A_{0n-2},
\label{Sn} 
\eeq

\noindent where the first term in Eq.~(\ref{Sn}) represents the
$n-$halo term, the second the contributions from $m$-halo terms, and
the last is the 1-halo term. The coefficients $\alpha_{nm}$ measure
how many of the terms contribute as $A_{1n-m-1}$, the other
contributions being subdominant. For example, in Eq.~(\ref{S4}) the
2-halo term has a total contribution of 7 terms, 4 of them contain 3
particles in one halo and 1 in the other, and 3 of them contain 2
particles in each. The factor $4/7$ is included to take into account
that the $3-1$ amplitude dominates over the $2-2$ amplitude. Note that
in these results we neglected all the contributions from the
non-linear biasing parameters in view of the consistency conditions,
Eq.~(\ref{biascon}). When neglecting halo-halo correlations,
Eq.(\ref{Sn}) reduces to those in Sheth (1996) in the limit that halos
are point-size objects ($u(\tk,y)=1$). 

For the perturbative values, we use (Bernardeau 1994)

\beq
S_3^{\rm PT}= \frac{34}{7}-\gamma,\ \ \ \ \ 
S_4^{\rm PT}= \frac{60712}{1323} -\frac{62}{3} \gamma + \frac{7}{3}
\gamma^2,
\eeq
\beq
S_5^{\rm PT}= \frac{200575880}{305613}-\frac{1847200}{3969} \gamma +
\frac{6940}{63} \gamma^2 - \frac{235}{27} \gamma^3,
\eeq

\noindent where for simplicity we neglect derivatives of $\gamma$ with
respect to scale, which is a good approximation for $R \la 20$ Mpc/h.

Before we compare these predictions for dark matter clustering with
numerical simulations, it is important to note that, within this
framework, there are many ingredients which can be adjusted to improve
agreement with simulations. Rather than exploring all possible
variations, we have chosen to always use the NFW halo profile and the
dependence of the concentration parameter on mass given earlier, and
only change the mass function between PS and ST; it turns out that
these two models usually bracket the results of numerical simulations.
Other choices are considered in Seljak (2000), Ma \& Fry (2000),
Cooray \& Hu (2000). The sensitivity of the results to these choices
reflects the underlying uncertainty in this type of calculation. As
numerical simulation results converge on the different ingredients,
however, the predictive power of this method will increase.

\section{Comparison with Numerical Simulations}

We have run two sets of N-body simulations using the adaptive P$^3$M
code Hydra (Couchman, Thomas, \& Pearce 1995). Both have $128^3$
particles and correspond to a $\Lambda$CDM model ($\Omega=0.3$,
$\Omega_\Lambda=0.7$) with $\sigma_8=0.9$. The first set contains 14
realizations of a box-size 100 Mpc/h, and softening length $l_{\rm
soft}=100$ kpc/h. The second set has 4 realizations of a box-size 300
Mpc/h, and $l_{\rm soft}=250$ kpc/h, which allows us to check for
finite volume effects. We have also studied the effects of changing
the softening, number of time-steps and $\Omega$ as described below.

\subsection{Bispectrum Measurements: Finite Volume Effects} 

\begin{figure}[t!]
\centering
\centerline{\epsfxsize=18. truecm \epsfysize=15. truecm 
\epsfbox{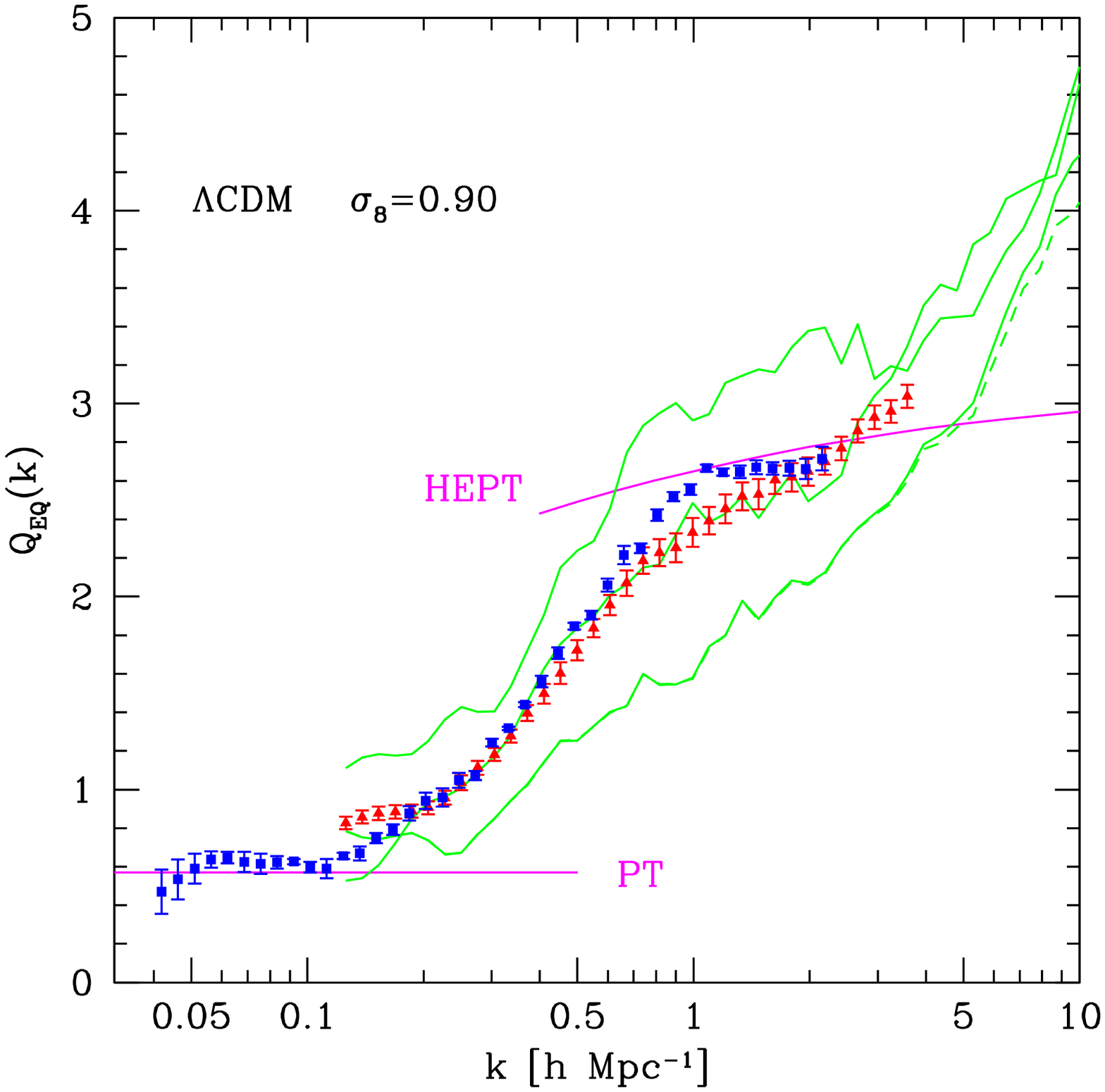}}
\caption{The reduced bispectrum $Q_{\rm eq}(k)$ for equilateral
configurations as a function of scale. The triangle symbols show the
average over 14 realizations of box size $L_{\rm box}=100$ Mpc/h; the
3 solid lines represent results for 3 individual realizations. The
dashed line denotes the same realization as the companion solid line
but ran with half the softening length (50 Kpc/h). The square symbols
denote the average over 4 realizations with $L_{\rm box}=300$
Mpc/h. The disagreement of the large and small box measurements is due
to finite-volume effects in the latter.}
\label{fig_qnbody}
\end{figure}

Figure~\ref{fig_qnbody} shows the results on the reduced bispectrum
for equilateral triangles as a function of scale. The square symbols
show the measurements from the ``large'' box ($L_{\rm box}=300$ Mpc/h)
whereas the triangle symbols show the measurements from the ``small''
box ($L_{\rm box}=100$ Mpc/h). Error bars are obtained from the
scatter among 4 and 14 realizations, respectively. The disagreement
between the results of the large and small boxes is a result of finite
volume effects; the bispectrum is much more sensitive to the presence
or absence of massive halos than the power spectrum (we will quantify
this below), so the smallness of the small box is important.  Note
that the total volume in the 14 small-box realizations only adds up to
about half of the volume of a single large-box realization. This
translates into a large scatter among realizations of the small box; 3
such realizations are shown as solid lines in Fig.~\ref{fig_qnbody}.
Note that not only the amplitude of $Q_{\rm eq}$ but also its
dependence on scale fluctuates significantly from realization to
realization, so one must interpret measurements made using only a
small number of small volume simulations very cautiously. A similar
situation holds for higher-order moments (e.g. Colombi, Bouchet \&
Hernquist 1996), as we shall see below.

As is well known, the distribution of higher-order statistics is
non-Gaussian with positive skewness (e.g. Szapudi \& Colombi 1996;
Szapudi et al. 1999); the mean value of a higher-order statistic is a
consequence of having a few large excursions above the mean, with most
values underestimating the mean. This is exactly what we see in the
small box realizations: most of them are closer to the bottom solid
line than the top one in Fig.~\ref{fig_qnbody}. Even fourteen
realizations of the small box are not enough to recover the correct
answer given by the large-box mean (square symbols).  In other words,
the skewness of the $Q$ distribution makes convergence towards the
true value much slower for the small-box simulations (Szapudi \&
Colombi 1996; see also Scoccimarro 2000 for the bispectrum case).
Notice that this is not bias in a statistical sense: given sufficient
number of realizations, the mean will always converge to the true
value; it is just that the convergence is slow.  It is also important
to note that when measuring $Q$ from multiple realizations, one should
always obtain the average of $B$ and the average of $P$ separately
from the realizations, and only at the end divide to obtain $Q$ (which
is what we have done in making Fig.~\ref{fig_qnbody}).  Otherwise, a
``ratio'' bias would result (Hui \& Gazta\~naga 1999), and the
skewness measurements from the small box simulations would be lower
than are shown in Fig.~\ref{fig_qnbody}. Such an estimator bias will
certainly affect measurements of $Q$ from, say, a single
realization of a size similar to our small box.


Figure~\ref{fig_qnbody} also shows the predictions of (tree-level) PT,
which agree very well with the large-box results at large scales, as
well as the predictions of hyperextended PT (HEPT; Scoccimarro \&
Frieman 1999), which has been proposed as a description of clustering
in the non-linear regime ($k \ga 1$ h/Mpc). The agreement with HEPT is
good up to the resolution scale of our simulations, which we estimate
as $k \approx 4$ h/Mpc.  It is not straightforward to assign a
resolution scale in Fourier space (i.e. it is not just $2\pi/l_{\rm
soft}$, since a given Fourier mode has contribution from a range of
scales. Two-body relaxation causes the two-point correlation function
to be underestimated at scales comparable to $l_{\rm soft}$, which in
turn implies an overestimate of the reduced bispectrum $Q$. To
illustrate this, we have run the same realization after halving
$l_{\rm soft}$ to $50$ Kpc/h (dashed line in Fig.~\ref{fig_qnbody});
this shows that beyond $k \approx 4$ h/Mpc the bispectrum results are
sensitive to the resolution, so we only plot the small-box results up
to this scale.  Similarly, we only show results for the large box up
to $k \approx 2$ h/Mpc. For our particle number, $l_{\rm soft}$ cannot
be pushed too much smaller than the values we used, else there would
be not enough particles in a cell of radius $l_{\rm soft}$ to satisfy
the fluid limit. We have also checked the sensitivity of our results
to changes in the number of time steps used in the N-body integrator
and found no difference. Changing the density parameter $\Omega$ leads
to extremely small differences in the reduced bispectrum $Q$, we thus
present results only for $\Lambda$CDM. This is true not only in the
weakly non-linear regime, but also in the non-linear regime, as
expected from the general nature of the $\Omega$ dependence in the
equations of motion (Scoccimarro et al. 1998). 

Recently Ma \& Fry (2000) claimed that the hierarchical ansatz is not
obeyed in N-body simulations. They based their claim on analysis of
one single realization---the equivalent to just one of our small
boxes.  Their measurement approximately follows the lower solid line
in Fig.~\ref{fig_qnbody}, which, as we have shown, is seriously
affected by finite volume effects (we will quantify these effects
shortly).  To reliably test the hierarchical ansatz at smaller scales
than probed here, one must resort to higher resolution simulations,
preferably keeping the box size as large as possible to avoid finite
volume effects. For example, a 300 Mpc/h box $512^3$ particle
simulation would be able to probe up to $k \approx 10$ h/Mpc reliably.

\subsection{Comparison with Predictions}

\begin{figure}[t!]
\centering
\centerline{\epsfxsize=18. truecm \epsfysize=15. truecm 
\epsfbox{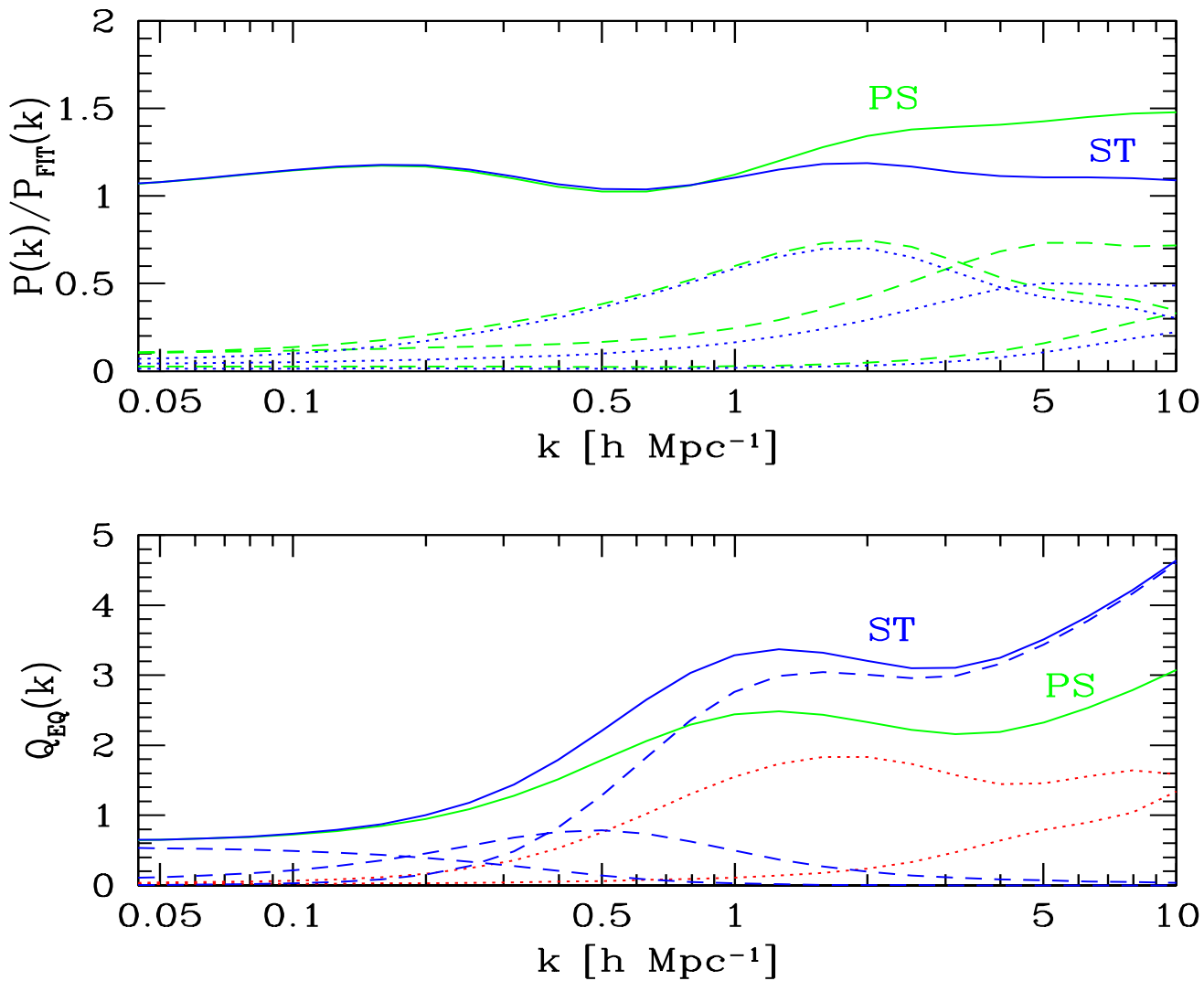}}
\caption{The top panel shows the ratio of power spectra for the PS and
ST mass function to the fitting formula for the non-linear power
spectrum of Peacock \& Dodds (1996) as a function of scale. The dashed
(dotted) lines show the contributions to the 1-halo term in the PS
(ST) case from halos of mass $10 < m/m_* < 100$, $1 < m/m_* < 10$ and
$0.1 < m/m_* < 1$ from left to right. Bottom panel shows the
predictions for the reduced equilateral bispectrum; the dashed lines
show the individual contributions (for the ST case) of 3-halo, 2-halo
and 1-halo terms, which dominate at large, intermediate and small
scales, respectively. The dotted lines show the contributions to the
1-halo term in $Q$ in the PS case from halos in the mass range $10 <
m/m_* < 100$ and $1 < m/m_* < 10$.}
\label{fig_pkqeq}
\end{figure}

The top panel in Fig.~\ref{fig_pkqeq} shows the ratio of the power
spectrum in our two models (PS and ST mass functions) to the power
spectrum fitting formula (Hamilton et al. 1991; Jain, Mo \& White
1995; Peacock \& Dodds 1996) as a function of scale $k$. The dashed
(dotted) lines show the contributions to the 1-halo term in the PS
(ST) case from halos having masses in the range $10<m/m_*<100$,
$1<m/m_*<10$, and $0.1<m/m_*<1$, from left to right ($m_* = 1.07\times
10^{13}M_{\sun}$/h). As the PS mass function has more halos than the
ST one when $m \la 40 m_*$, the 1-halo term is enhanced. Note the dip
in both predictions at $k \approx 0.5$ h/Mpc, where the amplitude of
1-halo and 2-halo terms are comparable. This is due to our treatment
of the 2-halo term; we approximate it by simply using linear PT. In
practice, non-linear corrections enhance this term at scales smaller
than the non-linear scale, $k \approx 0.3$~h/Mpc. However, when
including this term one must also take into account exclusion effects
(halos cannot be closer than their typical size), otherwise the power
spectrum at intermediate scales would be overestimated. Since
exclusion effects are non-trivial to compute (though Sheth \& Lemson
1999 suggest how one might do so), the simplest solution is to ignore
these effects, because they approximately cancel each other.  This is
a reasonable approximation because at the scales where exclusion
effects become important, 1-halo contributions dominate.

The bottom panel in Fig.~\ref{fig_pkqeq} shows the prediction of halo
models for the reduced bispectrum for equilateral configurations as a
function of scale (solid lines). For the ST case we also show the
partial contributions from 1-halo, 2-halo and 3-halo terms in dashed
lines, which dominate at small, intermediate and large scales,
respectively. For the PS case we show in dotted lines the
contributions to the 1-halo term in $Q$ when the bispectrum is
restricted to the mass range $10<m/m_*<100$ (which dominates at all
scales shown in the plot) and $1<m/m_*<10$. In this case, when taking
the ratio in Eq.~(\ref{Q}), we have used the full power spectrum. As
expected, comparing the two panels we see that at a given scale the
bispectrum is dominated by larger mass halos than the power spectrum. 
For the bispectrum at $k \approx 1$h/Mpc, this implies that
halos with $m>40 m_*$ contribute more significantly, thus leading to a
higher $Q$. At smaller scales, say $k \approx 10$h/Mpc, the PS mass
function has more halos of the relevant masses ($m<40 m_*$), so the
bispectrum is larger for PS than ST (by a factor slightly smaller than
the ratio of power spectra), thus the reduced bispectrum $Q$ is higher
again for the ST case. We have also varied the concentration parameter
to test the sensitivity of our results. Doubling the concentration
parameter (with the same mass dependence) leads to a significant
increase of the power spectrum at small scales (this consistent with 
Seljak 2000) and increases $Q$ by about 10\% at small scales. On the 
other hand, changing the scaling of the concentration parameter by a 
factor of two to $c(m)= 9 (m/m_*)^{-0.26}$, decreases (increases) the 
power spectrum at scales where $m/m_*>1$ ($m/m_*<1$) contributes, as 
expected. For the bispectrum, larger concentration leads to higher $Q$, 
although at a given scale larger masses contribute than for the power 
spectrum, so the effects are shifted in scale with respect to the 
power spectrum case.

\begin{figure}[t!]
\centering
\centerline{\epsfxsize=18. truecm \epsfysize=15. truecm 
\epsfbox{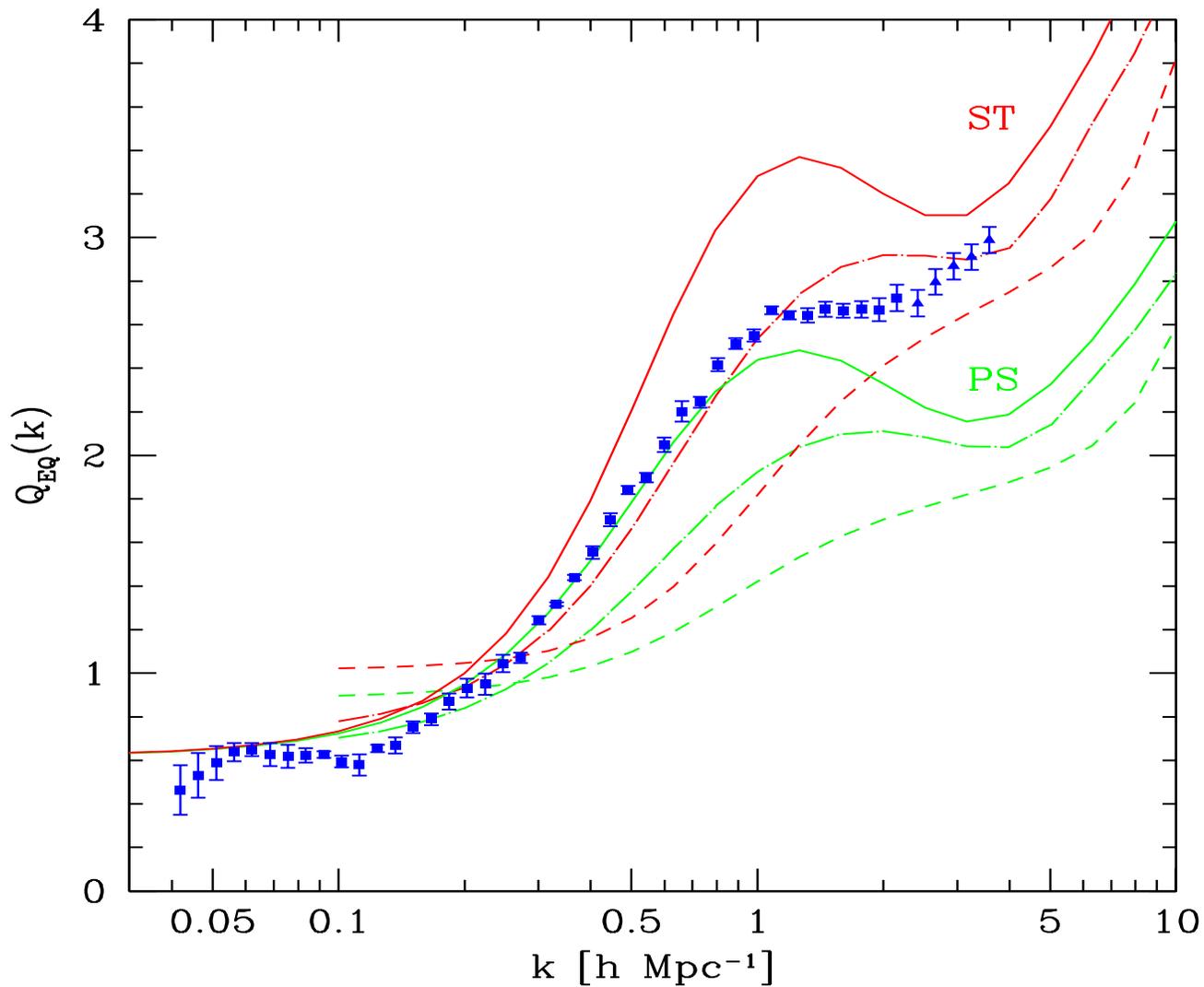}}
\caption{Comparison of the N-body results of
Fig.~\protect\ref{fig_qnbody} (points with error-bars) with the predictions of
Fig.~\protect\ref{fig_pkqeq} (solid lines). 
The dashed lines show the predictions of
halo models when there are no halos of mass larger than $m=10^{14}~
M_{\sun}$/h in the simulation volume, to illustrate finite volume
effects, for both PS (lower curve) and ST mass functions. 
The dot-dashed lines represent the predictions if halos of
mass larger than $m = 5.9 \times 10^{14} M_{\sun}/h$ (or $m = 6.8 \times 
10^{14} M_{\sun}/h$) are excluded for PS (or ST) mass functions.}
\label{fig_Qnb_halo}
\end{figure}

\begin{figure}[t!]
\centering
\centerline{\epsfxsize=18. truecm \epsfysize=15. truecm 
\epsfbox{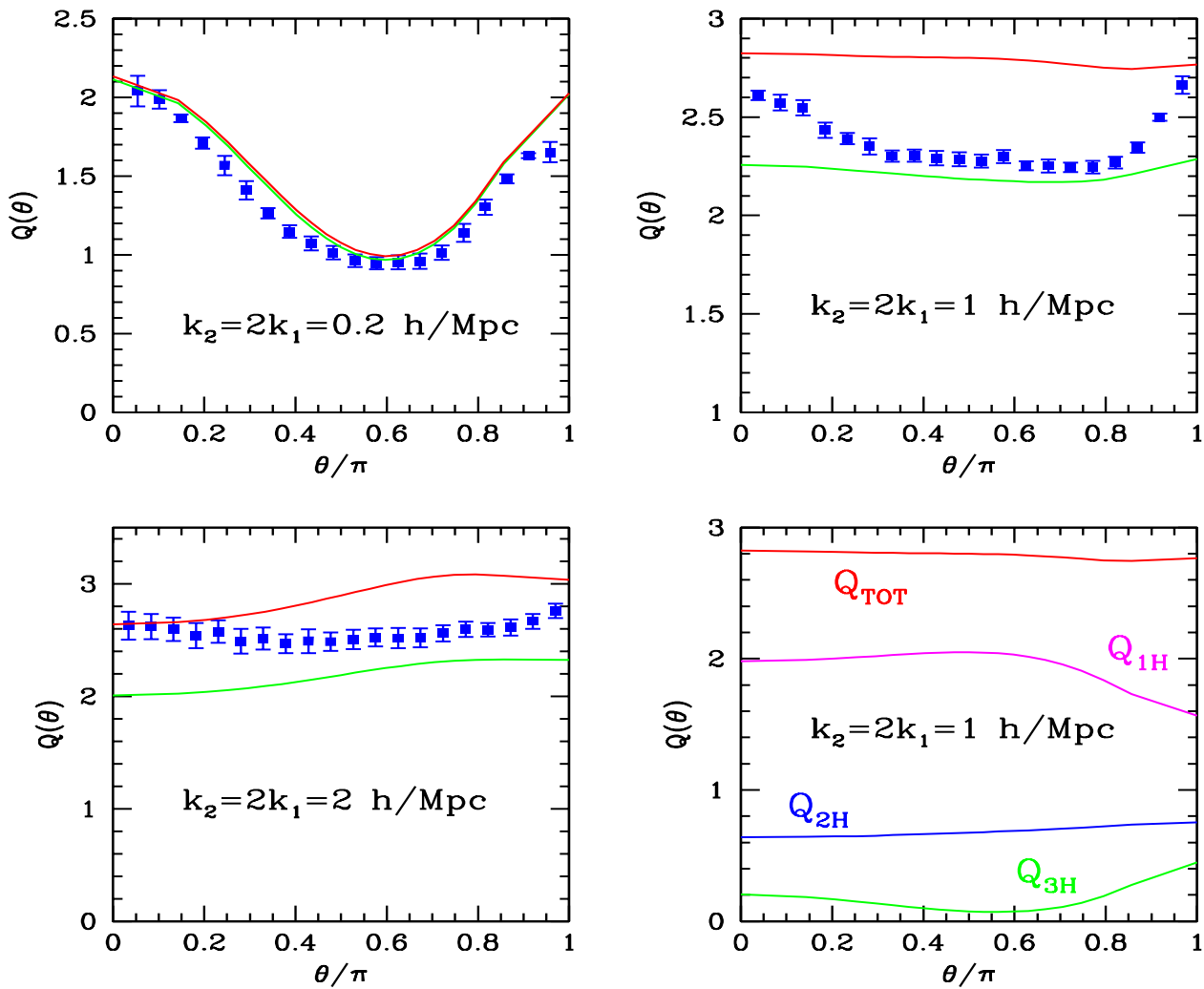}}
\caption{Comparison of the N-body results with the predictions of halo
models for the reduced bispectrum as a function of triangle
shape. Symbols and error bars are as in
Fig.~\protect\ref{fig_qnbody}. Solid lines show the predictions of ST
(top) and PS (bottom) mass functions.  The top panels and lower left
panel show different scales, whereas the lower right panel shows the
partial contributions to the total value, from 1-halo, 2-halo and 3-halo
terms for the ST case.}
\label{fig_Qnb_halo_config}
\end{figure}

Figure~\ref{fig_Qnb_halo} compares these results with the measurements
in the numerical simulations presented in Fig.~\ref{fig_qnbody}.  We
see that generally there is good agreement between predictions and the
simulations; the simulations seem to be roughly in between the PS and
ST predictions. At small scales, the halo models predict that $Q_{eq}$
increases rapidly with $k$; the limited resolution of our simulations
prevents us from testing this particular prediction reliably.  As we
discussed above, finite volume effects can be significant when dealing
with the bispectrum. To quantify this, we have calculated the halo
model predictions for cases when the maximum halo mass is set to
$m_{\rm max}= 5.9\times 10^{14} M_{\sun}$/h and $m_{\rm max}=
6.8\times 10^{14} M_{\sun}$/h for PS and ST respectively (dot-dashed
lines) and $m_{\rm max}= 10^{14} M_{\sun}$/h (dashed lines). These
values are those for which the mass functions would predict just one
halo with mass larger than $m_{\rm max}$ in a $(100 {\rm Mpc/h})^3$
volume; but since these are cumulative numbers and both mass functions
actually overestimate the number of halos when compared to simulations
(more so PS), a smaller number, such as $m_{\rm max}= 10^{14}
M_{\sun}$/h is perhaps a more reasonable cutoff.  In any case, we see
that the predictions change significantly; in particular, introducing
such a cutoff makes the scale dependence of $Q$ much more like that
seen in most of the realizations of the small box (bottom solid line
in Fig.~\ref{fig_qnbody}).

One key element in the halo model is that we are using the {\em
spherical average} (rather than the actual shapes) of halo
profiles. On the other hand, it is well known that halos found in
N-body simulations are not spherical, but rather triaxial (Barnes \&
Efstathiou 1987; Frenk et al. 1988; Zurek, Quinn \& Salmon 1988). The
bispectrum is the lowest-order statistic which is sensitive to the
shapes of structures, so one expects to find differences for the
bispectrum as a function of triangle shape at small scales where halo
profiles (1-halo terms) determine correlation functions.
Figure~\ref{fig_Qnb_halo_config} shows such a comparison at different
scales, for triangles where $k_2=2k_1$, as a function of angle
$\theta$ between $\k_1$ and $\k_2$. The top left panel shows that, at
large scales, the bispectrum agrees reasonably well with simulations;
this is of course by construction, since non-linear PT holds. At
smaller scales (top right panel), however, the predictions become
independent of triangle shape at scales where there is still
noticeable configuration dependence. In fact, this is understood from
the bottom right panel which shows the partial contributions for the
ST case. We see that the 1-halo term (which is determined by the halo
profiles) has the {\em opposite} configuration dependence than the
3-halo term, which comes from non-linear PT.

The fact that $Q_{1h}$ is convex can be understood from the spherical
approximation.  If halos were exactly spherical and $Q$ were
scale-independent, then one would expect the maximum of $Q$ to occur
for equilateral configurations.  When $k_2=2k_1$ the closest
configurations to equilaterals are isosceles triangles with
$\theta\approx 0.6 \pi$. Aside from an overall slight scale dependence
($\theta=0$ configurations are somewhat more non-linear than
$\theta=\pi$), we see that this is indeed the case. Notice also that,
if the contribution $Q_{1h}$ were flat, the residual configuration
coming from $Q_{3h}$ would be enough to produce agreement with the
N-body simulations.  At even smaller scales (bottom left panel), the
numerical simulation results become approximately flat, but the halo
models predict a convex configuration dependence due to the fact that
$Q_{1h}$ dominates.  From these results we conclude that although
halo models predict bispectrum amplitudes which are in reasonable
agreement with simulations, the configuration dependence is in
qualitative disagreement with simulations.  Of course, if we knew the
actual halo shapes, then they could be incorporated into the models
(at the expense of complicating the calculations!).

\begin{figure}[t!]
\centering
\centerline{\epsfxsize=18. truecm \epsfysize=15. truecm 
\epsfbox{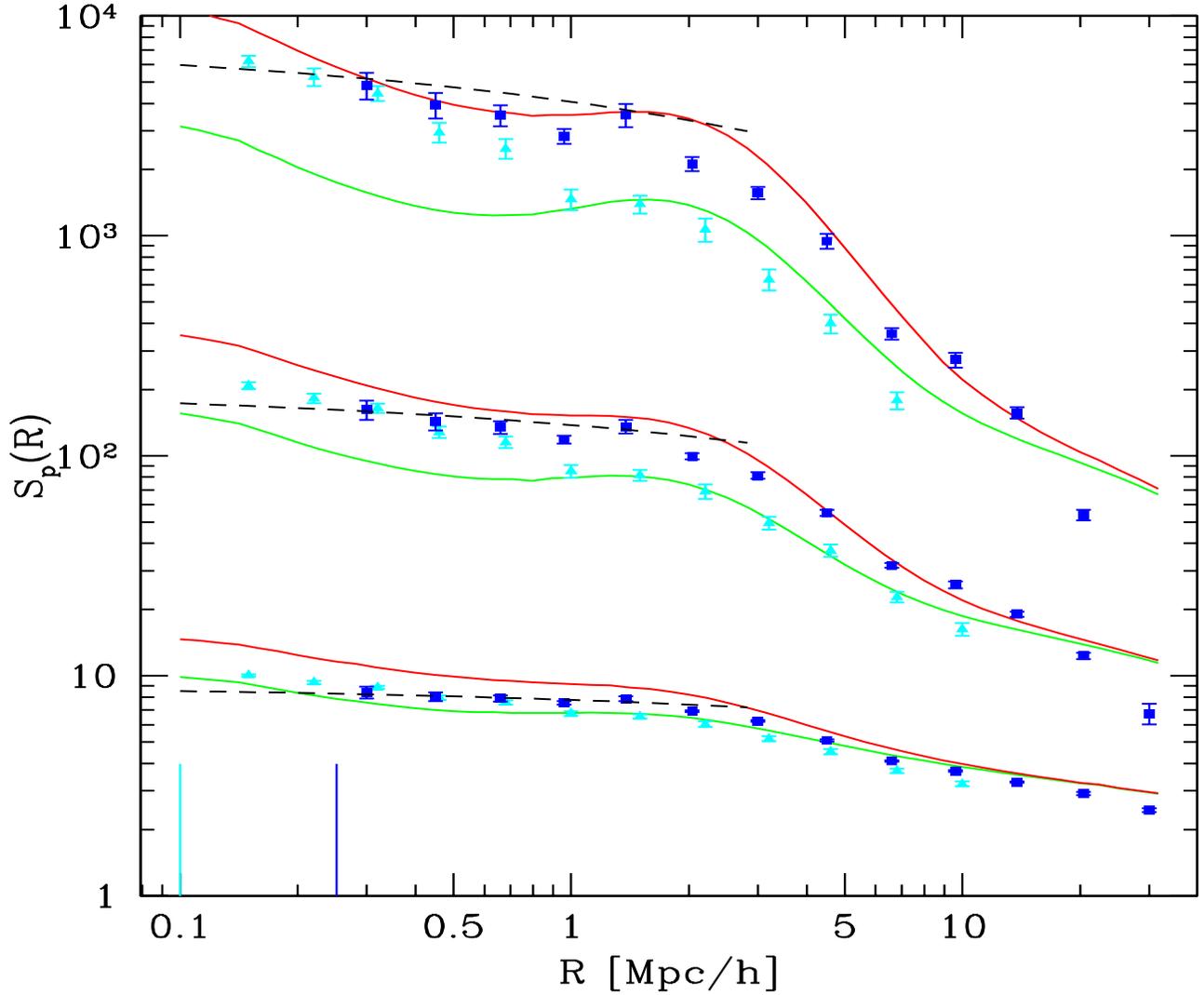}}
\caption{$S_p$ parameters ($p=3,4,5$ from bottom to top) as a function
of smoothing scale $R$. Symbols and error bars are as in
Fig.~\protect\ref{fig_qnbody}. Solid lines show the predictions of ST
(upper) and PS (lower) mass functions. Dashed lines show the
asymptotic behavior predicted by HEPT. Again, the disagreement of the
small volume simulations (triangles) with those of bigger volume
(squares) is a direct consequence of finite volume effects in the
former. The vertical lines denote the softening length of the two sets
of simulations. }
\label{fig_Sp_NB}
\end{figure}

Finally, in Fig.~\ref{fig_Sp_NB}, we compare counts-in-cells
measurements of the higher-order moments of the smoothed density field
in our simulations with the predictions of halo models. Symbols and
error bars are as in Fig.\ref{fig_qnbody}: the top (bottom) solid line
in each case corresponds to the ST (PS) prediction. The dashed lines
show the predictions of HEPT (Scoccimarro \& Frieman 1999), and the
vertical lines show the softening scale of the large and small
box. The disagreement of the average of 14 small-box measurements with
the large-box average is, again, a manifestation of finite volume
effects.  As expected, the difference becomes increasingly important
for the higher order moments. Despite the many approximations made in
the calculations of $S_p$ parameters in halo models, the agreement
with simulations is quite good. We also see a very good agreement with
the HEPT predictions, and that the scales where halo models predict a
strong scale dependence are beyond the limits of our resolution. This
also confirms that our prescription for the resolution limit in
Fourier space was reasonably accurate.  Thus, contrary to Ma \& Fry
(2000), we conclude that higher-resolution simulations in bigger boxes
are essential if one is to test models of the higher-order
correlations reliably.

\section{Galaxy Clustering} 

We now turn to a discussion of how to use the halo models described
above to predict the clustering of galaxies.  Our treatment follows
ideas present in the semianalytic galaxy formation models (Benson et
al. 2000; Kauffmann et al. 1999) and has been also explored by Seljak
(2000) and Peacock \& Smith (2000) for the case of the power spectrum.

\subsection{Galaxy Correlation Functions} 

To describe {\em galaxy} clustering, we need to know the distribution, 
the mean and the higher-order moments, of the number of galaxies which 
can inhabit a halo of mass $m$. In addition, we need to know the 
spatial distribution of galaxies within their parent halo.  
To illustrate our model predictions, in what follows we will assume 
that the galaxies follow the dark matter profile (we will discuss 
what happens if we change this requirement shortly).  This implies
that Eq.~(\ref{pkm}) for galaxies reads

\beqa
\bar{n}_g^2 P_g(k) &=& (2\pi)^3 \int n(m) \lexp N_{\rm gal}^2(m) \rexp   
dm |u_m(\k)|^2 + \nonumber \\ & & 
(2\pi)^6 \int u_{m_1}(k) n(m_1) \lexp N_{\rm gal}(m_1)\rexp  dm_1 
\int u_{m_2}(k) 
n(m_2) \lexp N_{\rm gal}(m_2)\rexp  dm_2 P(k;m_1,m_2),
\label{pkg}
\eeqa

\noindent where the mean number density of galaxies is

\beq
\bar{n}_g = \int n(m) \lexp N_{\rm gal}(m)\rexp  dm.
\eeq

\noindent Thus, knowledge of the number of galaxies per halo moments
$\lexp N_{\rm gal}^n(m) \rexp$ as a function of halo mass gives a
complete description of the galaxy clustering statistics within this
framework. Note that when the mean number of galaxies per halo drops
below unity, one must use $u_m(\k)=1$, the point-size halo limit,
since in this case there is at most a single galaxy (which we assume
to be at the center of the halo).

The results for galaxy power spectrum and bispectrum follow those of
the dark matter in Eqs.(\ref{power}-\ref{bisp}), after changing 
$M_{ij}$ in Eq.~(\ref{Iij}) to $G_{ij}$, where 

\beq
G_{ij}(\tk_1,\ldots,\tk_j) \equiv \frac{1}{\bar{n}_g^{j}} 
\int \frac{dy}{y} \tilde n(y) b_i(y) [u(\tk_1,y)\ldots u(\tk_j,y)]
\Big(\frac{R_*^3 y^3}{6\pi^2} \Big)^{j-1} \frac{\lexp N_{\rm gal}^j
\rexpc}{m^j}.
\label{Gij}
\eeq

\noindent Note that in the large-scale limit, the galaxy bias
parameters are 

\beq
b_i = G_{i1} \approx  \frac{1}{\bar{n}_g} \int \frac{dy}{y} \tilde n(y)
b_i(y) \frac{\lexp N_{\rm gal} \rexp}{m}.
\label{beff}
\eeq

\noindent Similarly, the galaxy one-point connected moments satisfy

\beq
\sigma_g^2=(\sigma^2_L)_{\rm gal}+B_{00},\ \ \ \ \ 
\lexp \d_g^3 \rexpc= (S_3^{\rm PT})_{\rm gal} (\sigma^4_L)_{\rm gal} +
3 (\sigma^2_L)_{\rm gal} B_{10}+B_{01},
\label{vS3g} 
\eeq

\beq
\lexp \d_g^4 \rexpc= (S_4^{\rm PT})_{\rm gal} (\sigma^6_L)_{\rm gal} +
2 S_3^{\rm hh} (\sigma^4_L)_{\rm gal} B_{10}+4
(\sigma^2_L)_{\rm gal} B_{11}+B_{02},
\label{S4g} 
\eeq

\noindent and

\beq \lexp \d_g^5 \rexpc = (S_5^{\rm PT})_{\rm gal} (\sigma^8_L)_{\rm
gal} + \frac{5}{8} S_4^{\rm hh} (\sigma^6_L)_{\rm gal} B_{10} + 15
S_3^{\rm hh} (\sigma^4_L)_{\rm gal} B_{11} + 5 (\sigma^2_L)_{\rm gal}
B_{12} +B_{03},
\label{S5g} 
\eeq

\noindent where

\beq
B_{ij}(R) \equiv \frac{1}{b_i \bar{n}_g^{j+2}}
\int \frac{dy}{y} \tilde n(y)\ b_i(y)\ y^{3(j+1)}\ 
[\overline{u}(R,y)]^j\  \overline{u^2}(R,y) \frac{\lexp N_{\rm
gal}^{j+2} \rexp}{m^{j+2}},
\label{Bij}
\eeq

\noindent and the perturbative moments are given by their local bias
counterparts (Fry \& Gazta\~naga 1993)

\beq
(\sigma^2_L)_{\rm gal} = b_1^2 \sigma^2_L,\ \ \ \ \ 
(S_3^{\rm PT})_{\rm gal} = \frac{1}{b_1} \Big(S_3^{\rm
PT}+3c_2\Big),\ \ \ \ \
(S_4^{\rm PT})_{\rm gal} = \frac{1}{b_1^2} \Big(S_4^{\rm
PT}+12c_2 S_3^{\rm
PT}+4c_3+12c_2^2\Big),
\label{S3S4g}
\eeq

\noindent and

\beq
(S_5^{\rm PT})_{\rm gal} = \frac{1}{b_1^3} \Big(S_5^{\rm
PT}+20 c_2 S_4^{\rm PT}+15 c_2(S_3^{\rm PT})^2+(30c_3+
120c_2^2)S_3^{\rm PT}+  5c_4+60c_2 c_3+60c_2^3\Big),
\label{S5g2}
\eeq

\noindent where $c_i\equiv b_i/b_1$ and $b_i$ are the effective bias
parameters in Eq.~(\ref{beff}). The halo-halo skewness and kurtosis
are given by these expressions upon replacing $c_i$ by $c_i
B_{i0}/B_{10}$.

\subsection{Galaxies} 

As a first example, we use the results of the semi-analytic models of
Kauffmann et al. (1999); these N-body simulations, halo and galaxy
catalogues are publically available.  Sheth \& Diaferio (2000) show
that in these catalogues, the mean number of galaxies $N_{\rm gal}$
per halo of mass $m$ are well fit by

\beq
\lexp N_{\rm gal} \rexp = \lexp  N_B+N_R \rexp,\ \ \ \ \ 
\lexp N_B \rexp=0.7 (m/m_B)^{\alpha_B},\ \ \ \ \ 
\lexp N_R \rexp=(m/m_R)^{\alpha_R},
\label{SD}
\eeq

\noindent where $N_B$ and $N_R$ represent the number of blue and red
galaxies per halo of mass $m$, and $\alpha_B=0$ for $10^{11}
M_{\sun}/h \leq m \leq m_B$, $\alpha_B=0.8$ for $m > m_B$,
$m_B=4\times 10^{12} M_{\sun}/h$, $\alpha_R=0.9$, and $m_R=2.5\times
10^{12} M_{\sun}/h$ (no lower mass cut-off for $R$). The physical
basis for this relation is as follows. At large masses, the gas
cooling time becomes larger than the Hubble time, so galaxy formation
is suppressed in large-mass halos, therefore $\lexp N_{\rm gal} \rexp$
increases less rapidly than the mass. In small-mass halos, however,
effects such as supernova winds can blow away the gas from halos, also
suppressing galaxy formation, thus the cutoff at small masses.

To calculate the power spectrum and higher-order statistics we also
need the second and higher-order moments of $N_{\rm gal}$. The
second moment is also obtained from the semi-analytic models, and
obeys

\beq
\lexp N_{\rm gal} (N_{\rm gal}-1) \rexp \equiv \alpha^2(m) \lexp
N_{\rm gal} \rexp^2,
\label{nnm1}
\eeq
 
\noindent where the function $\alpha(m)$ quantifies deviations from
Poisson statistics $\alpha(m) \approx \log \sqrt{m/m_{11}}$ ($m_{11}
\equiv 10^{11} M_{\sun}/h$) for $m<10^{13} M_{\sun}/h$ and
$\alpha(m)=1$ otherwise; that is, for large masses the scatter about
the mean number of galaxies is Poisson, whereas for small masses it is
sub-Poisson.

To model the higher-order moments we will assume that the number of
galaxies in a halo of mass $m$ follows a binomial distribution: 

\beq
p(N_{\rm gal}=n|m) = {{\cal N}_m\choose n} p_m^n (1-p_m)^{{\cal
N}_m-n}.  
\eeq 

The binomial distribution is characterized by two parameters, ${\cal
N}_m$ and $p_m$, which we set by requiring that the first two moments
of the distribution equal those from the semianalytics.  Specifically,
the first and second factorial moments are ${\cal N}_mp_m$ and ${\cal
N}_mp_m({\cal N}_mp_m-p_m)$, and we require that they equal $\lexp
N_{\rm gal} \rexp$ and $\lexp N_{\rm gal}(N_{\rm gal}-1)\rexp$,
respectively.  One can think of ${\cal N}_m$ as the maximum number of
galaxies which can be formed with the available mass $m$, and of $p_m$
as the probability of actually forming a galaxy. For a constant ${\cal
N}_mp_m$, the small $p_m$ limit gives a Poisson distribution. ${\cal
N}_m$ is an increasing function of mass, whereas $p_m$ peaks at $m
\approx 10^{12} M_{\sun}/h$ with $p_m \approx 0.8$. The higher-order
factorial moments are completely determined once the first two moments
have been specified; they obey

\beq
\lexp N_{\rm gal} (N_{\rm gal}-1) \ldots (N_{\rm gal}-j) \rexp =
 \alpha^2 (2\alpha^2-1) \ldots (j\alpha^2-j+1)
\lexp N_{\rm gal} \rexp^{j+1}.
\label{binom}
\eeq

\noindent In this model, all the moments become Poisson-like at the
same mass scale, i.e. when $\alpha(m)=1$, all the factorial moments
become Poisson, $\lexp N_{\rm gal}(N_{\rm gal}-1)\ldots (N_{\rm
gal}-j) \rexp =\lexp N_{\rm gal} \rexp^{j+1}$.  However, at small
scales, where small halos contribute, the galaxy counts per halo are
significantly sub-Poisson.  Our binomial model provides a simple way
of accounting for this.  To correct the power spectrum and bispectrum
for shot noise we use the same form as in the Poisson case (Peebles
1980):

\beq
P^{\rm c}(k) =P(k) -\varepsilon,\ \ \ \ \
B_{123}^{\rm c}=B_{123}-\varepsilon(P_1^{\rm c}+P_2^{\rm
c}+P_3^{\rm c})- \varepsilon^2,
\eeq

\noindent but where the parameter $\varepsilon$ is set to $\varepsilon
\equiv P(k \rightarrow \infty)$, to avoid making the corrected power
spectrum negative at small scales. In the Poisson case,
$\varepsilon^{-1} = (2\pi)^3 \bar{n}_g$, we find that in our
prescription $\varepsilon$ can be smaller than the Poisson value by
almost a factor of two. Although this model is somewhat arbitrary, our
results are insensitive to shot noise substraction for $k \la 5$h/Mpc
and smoothing scales $R \ga 1$Mpc/h.

\begin{figure}[t!]
\centering
\centerline{\epsfxsize=18. truecm \epsfysize=15. truecm 
\epsfbox{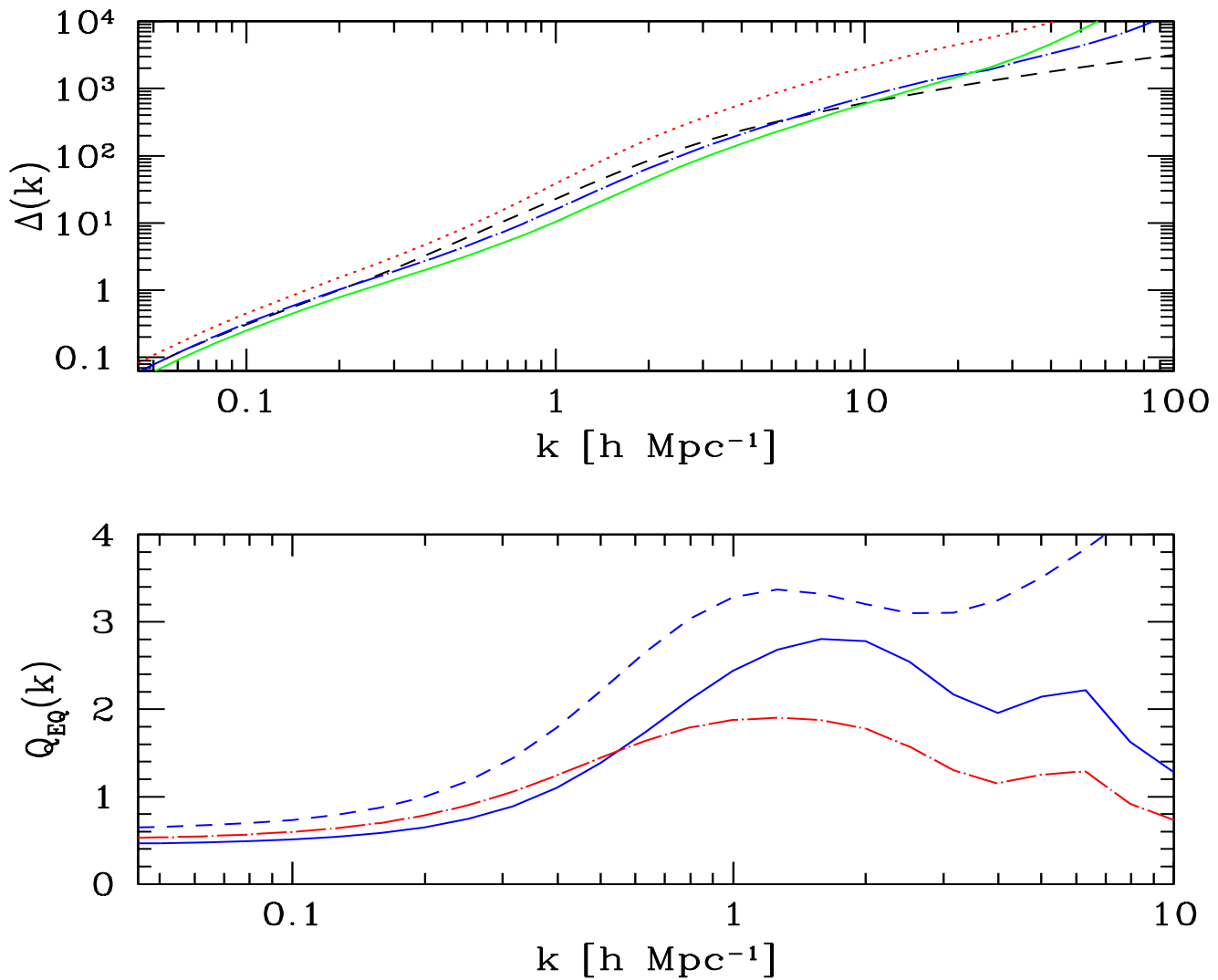}}
\caption{The top panel shows the galaxy power spectrum ($\Delta (k) =
4 \pi k^3 P(k)$) as a function of scale predicted by halo models for
ST (dot-dashed) and PS (solid) mass functions, using the $N_{\rm gal}
(m)$ given in Eq. (\ref{SD}).  The dashed line shows the mass power
spectrum predicted by the fitting formula for the ST case. The dotted
line shows the power for the ST case if the lower mass cut-off is
changed from $m_{\rm cut} = 10^{11} M_{\sun}/h$ to $m_{\rm cut} =
10^{11.5} M_{\sun}/h$.  The bottom panel shows the reduced galaxy
bispectrum for equilateral triangles as a function of scale for
$m_{\rm cut} = 10^{11} M_{\sun}/h$ (solid), $m_{\rm cut} = 10^{11.5}
M_{\sun}/h$ (dot-dashed), and for the mass (dashed). These assume the
ST mass function.}
\label{fig_PkQeq_gal}
\end{figure}

Figure~\ref{fig_PkQeq_gal} shows the resulting galaxy correlation
functions. The top panel shows the predictions of the ST (dot-dashed)
and PS (solid) mass functions for galaxies using the relations given
above. For comparison, we also show the mass power spectrum predicted
by the fitting formula of Peacock \& Dodds (1996). Note how the galaxy
power spectrum is well approximated by a power law, even though the
dark matter spectrum is not. This is due to the fact that galaxy
formation is inefficient in massive halos; this suppresses the 1-halo
term compared to the dark matter case. In addition, at small scales
the sub-Poisson statistics of galaxy counts per halo suppresses the
galaxy spectrum relative to the dark matter. The dotted line in the
top panel shows how the predictions change for the ST model when the
low-mass cutoff in the $N_{\rm gal}(m)$ relation is raised from
$m_{\rm cut}=10^{11} M_{\sun}$/h to $m_{\rm cut}=10^{11.5}
M_{\sun}$/h.  At large scales, suppressing low-mass halos leads to an
increase of the bias factor from $b_1=0.86$ to $b_1=1.12$.  At small
scales, since these low-masses do not contribute significantly, the
overall amplification of the 1-halo term is due to the lower galaxy
number density, which decreases by almost a factor of two. Therefore,
our galaxy clustering predictions are quite sensitive to the low-mass
cutoff for galaxy formation.

The bottom panel in Fig.~\ref{fig_PkQeq_gal} shows the galaxy reduced
bispectrum $Q_{\rm gal}$ for equilateral configurations as a function
of scale (solid lines), compared to the dark matter value (dashed) for
the ST mass function. At large scales, the halo models predict a
negative quadratic bias, which suppresses the bispectrum. At smaller
scales, both the power spectrum and bispectrum are suppressed, but at
a given scale the bispectrum is sensitive to larger mass halos than
the power spectrum, so $Q_{\rm gal}$ rises more gradually than $Q_{\rm
dm}$.  At small scales $k \ga 1$h/Mpc, the different mass weighing of
the galaxy bispectrum leads to a suppression of the scale dependence
of $Q$.  The dot-dashed curve shows again the sensitivity to the
low-mass cutoff ($m_{\rm cut}=10^{11.5} M_{\sun}$/h); as expected, the
distribution at small scales is more biased. To be more quantitative,
let's note that at small scales, where 1-halo terms dominate, the
scaling of the $p$-point spectrum is

\beq T_p(k) \sim \int n(m)\ m^p\ [u_m(k)]^p\ \frac{\lexp N_{\rm gal}^p
\rexp}{m^p}\ dm, \eeq

\noindent and at small masses $m \ll m_*$, $n(m) \sim m^{-2}
\nu^a(m)$, $c(m) \sim m^{-b}$ and $u_m(k) \sim [\sin \kappa\ {\rm
Sic}(\kappa)-\cos \kappa\ {\rm Ci}(\kappa)]/\ln c $, where ${\rm
Sic}(\kappa) \equiv \pi/2-{\rm Si}(\kappa)$. This expression for the
profile follows when $c \gg 1$, at large $\kappa$, $u_m(k) \sim
(\kappa \ln c)^{-1}$. Furthermore, if $\nu(m) \sim m^{(n+3)/6}$ as in
the scale-free case, and $\lexp N_{\rm gal}^p \rexp \sim
m^{p(1-\epsilon)}$, changing variable from $m$ to $\kappa$, we have
\beq
T_p(k) \sim k^{[6(1-p+p \epsilon)-(n+3)a]/[2(3b+1)]},
\eeq

\noindent where we have neglected the weak logarithmic dependence of
the profile on the concentration (which leads to additional suppresion
at small scales) and used that the integral over $\kappa$
converges. This means that the reduced spectra $Q_p(k) \sim
T_p(k)/[P(k)]^{p-1}$ scale as $Q_p(k) \sim k^{\gamma_p}$ with 

\beq
\gamma_p = (p-2) \frac{a(n+3)-6\epsilon}{2(3b+1)}.
\eeq

\noindent This generalizes the derivation in Ma \& Fry (2000b) for
galaxies. The validity of the hierarchical ansatz, $\gamma_p \approx
0$, thus depends on the low-mass behavior of the concentration
parameter, mass function and $N_{\rm gal}$. For scale-free initial
conditions, the validity of the hierarchical ansatz for the mass
correlation functions is linked to that of stable clustering (Peebles
1980), although for higher-order correlation functions stable
clustering takes a different form than the usual two-point requirement
that pairwise peculiar velocities cancel the Hubble flow (Jain
1997). We see that for $p=3$ (the bispectrum), $a=0.4$ (ST mass
function), $\epsilon=0.1$, $n_{\rm eff} \approx -2.5$ ($k=3$ h/Mpc),
and $b= 0.13$, $\gamma_3 \approx -0.3$, which agrees approximately
with the behavior of $Q_{\rm gal}$ in Fig.~\ref{fig_PkQeq_gal}. This
is only qualitative, due to the many approximations involved. On the
other hand, it captures the general behavior of mass weighing, for
$b>0$, supressing contributions from massive halos ($\epsilon>0$)
preferentially places galaxies in more concentrated halos, thus at a
given $k$ one is probing outer regions of the NFW profile compared to
the mass case. Since in the outer regions $u(r) \sim r^{-3}$, $Q_{\rm
gal}$ is less scale dependent than $Q$.

\begin{figure}[t!]
\centering
\centerline{\epsfxsize=18. truecm \epsfysize=15. truecm 
\epsfbox{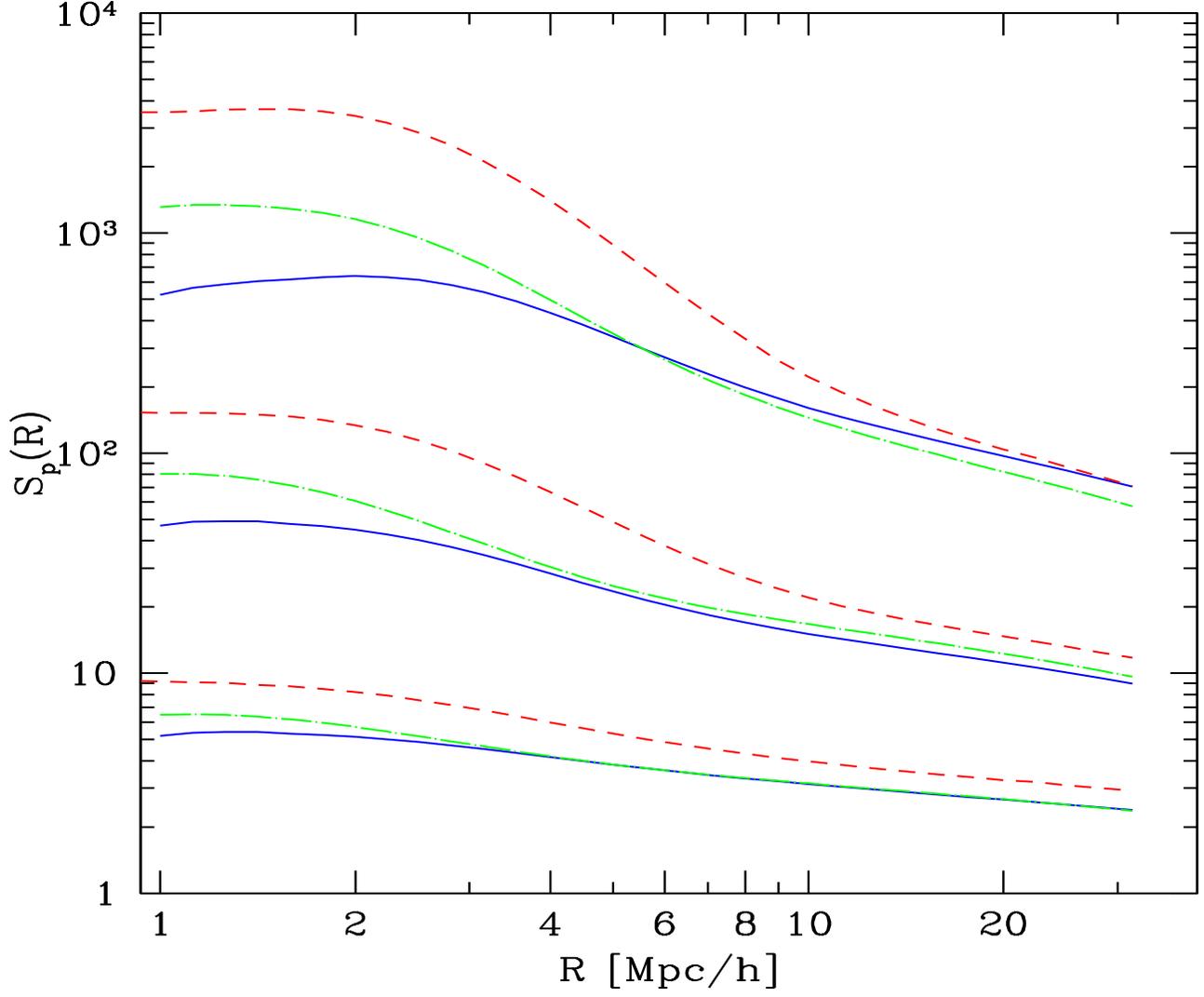}}
\caption{$S_p$ parameters ($p=3,4,5$ from bottom to top) as a function
of smoothing scale $R$ for the mass (dashed), galaxies as in
Eq.(\protect\ref{SD}) with low-mass cutoff at $m_{\rm cut}=10^{11}
M_{\sun}$/h (dot-dashed) and galaxies with $m_{\rm cut}=10^{11.5}
M_{\sun}$/h (solid).}
\label{fig_Sp_gal}
\end{figure}

Figure~\ref{fig_Sp_gal} shows the $S_p$ parameters as a function of
smoothing scale $R$ for the mass (dashed; same as ST in
Fig~\ref{fig_Sp_NB}), the galaxies with $m_{\rm
cut}=10^{11}M_{\sun}$/h (dot-dashed) and the galaxies with $m_{\rm
cut}=10^{11.5}M_{\sun}$/h (solid).  Although we see the same general
behavior as in the bispectrum case, it is interesting that the galaxy
$S_p$ parameters are smaller than the dark matter ones at small
scales, which is similar to the trend seen in comparisons of dark
matter predictions with real galaxy catalogs.  Both galaxy plots
assume that the galaxy per halo moments obey Eq.~(\ref{binom}). We
have repeated the calculation assuming Poisson statistics
($\alpha(m)=1$) and found that the results are the same for scales $R
\geq 5$ Mpc/h, and that at $R=1$ Mpc/h the Poisson values are about
15\% below those shown in Fig.~\ref{fig_Sp_gal}.  The sensitivity of
the clustering to the details of the relation of the number of
galaxies as a function of halo mass can be used to probe aspects of
galaxy formation, as we now discuss.

\section{Comparison with APM Survey: Constraints on Galaxy Formation}  

Measurements of two-point and higher order moments of the galaxy field
in the APM survey (Maddox et al. 1990) provide important constraints
on models of galaxy clustering. Here we use the measurements of
counts-in-cells, deprojected into three dimensions, by Gazta\~naga
(1994,1995) to constrain the $N_{\rm gal}(m)$ relation.

As discussed above, galaxy clustering at large scales is given by 
the standard local-bias model, with bias parameters obtained from
Eq.(\ref{beff}). To constrain the $N_{\rm gal}(m)$ relation,
we use the parametrized form

\beq
\lexp N_{\rm gal} \rexp = (m/m_0)^{a_1},\ \ \ \ \ 
m_{\rm cut} \leq m \leq m_0,\ \ \ \ \ \ \ \ \ \ 
\lexp N_{\rm gal} \rexp = (m/m_0)^{a_2},\ \ \ \ \ 
m \geq m_0,
\label{Ngcons}
\eeq

\noindent with $\lexp N_{\rm gal} \rexp=0$ for $m < m_{\rm cut}$ and
the second moment obeys Eq.~(\ref{nnm1}) with

\beq
\alpha(m)=\frac{\log(m/m_{\rm cut})}{\log(m_0/m_{\rm cut})},
\eeq

\noindent for $m \leq m_0$ and $\alpha(m)=1$ for $m \geq m_0$. For the
higher-order moments we adopt the binomial model in
Eq.(\ref{binom}). Requiring that $p_m$ in the binomial model to be
positive implies that $a_1 \geq 0$. Note that in Eq.~(\ref{Ngcons}) we
have set $\lexp N_{\rm gal} \rexp =1$ at $m=m_0$ since clustering
statistics do not depend on the overall number density of galaxies;
however, constraints such as the luminosity function and the
Tully-Fisher relation are sensitive to the overall amplitude of the
$N_{\rm gal}(m)$ relation.

For the measured values in APM at large scales, we use that
$\sigma_g^2=0.167 \pm 0.021$ at $R=20$ Mpc/h, and for the higher-order
moments we use $(S_3)_{\rm gal}=3.3\pm 0.5$ and $(S_4)_{\rm gal}=15
\pm 4$ at $R=20$ Mpc/h. The latter are fiducial values obtained by
averaging over the large scale ($R>10$) skewness and kurtosis, rather
than the value at a particular scale. The skewness at large scales is
rather flat so averaging is reasonable (see Fig.~\ref{fig_Sp_APM}),
for the kurtosis the large scale limit is not so well defined, but in
practice this constraint is not very important because it comes with
large error bars.

\begin{figure}[t!]
\centering
\centerline{\epsfxsize=18. truecm \epsfysize=15. truecm 
\epsfbox{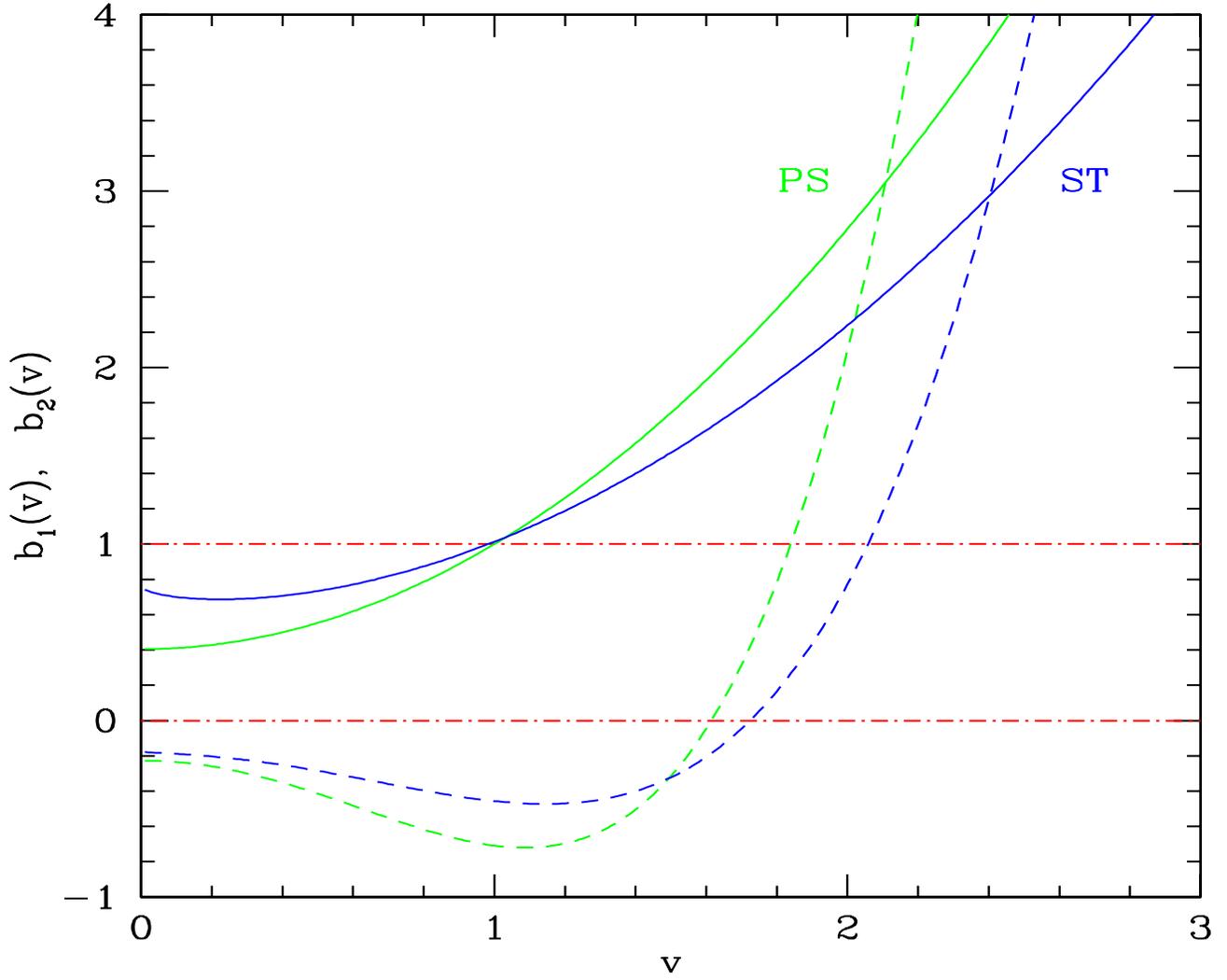}}
\caption{Halo linear ($b_1$, solid lines) and non-linear ($b_2$,
dashed lines) bias parameters as a function of threshold $\nu
=\d_c/\sigma(m)$ for the PS and ST mass functions. The dot-dashed
lines denote $b_1=1$ and $b_2=0$ for comparison.}
\label{fig_bias}
\end{figure}

The constraint on the variance at large scales sets the linear bias
parameter, $1 \la b_1 \la 1.15$, whereas the skewness also depends on
the non-linear bias $b_2$. In halo models, $b_1$ and $b_2$ are not
independent---for a given mass function they have a specific relation
as a function of halo mass, Eqs.~(\ref{b1}-\ref{b2}), shown in
Fig.\ref{fig_bias} in terms of the threshold parameter
$\nu=\d_c/\sigma(m)$.  As a result of this relation, it turns out to
be non-trivial to satisfy both $\sigma_g^2$ and $(S_3)_{\rm gal}$
constraints simultaneously. Essentially, since galaxies in cluster
normalized $\Lambda$CDM are constrained by the APM variance to be
almost unbiased at large scales ($b_1 \approx 1$), a high skewness (as
high as the mass skewness, shown as dashed lines in
Fig.~\ref{fig_Sp_APM}) requires that $b_2 \approx 0$, which is not
easy to obtain if galaxy formation is inefficient at small and large
halo masses (the latter is required to suppress the 1-halo term
contribution to the variance and match the observations at small
scales). To quantify constraints on the $N_{\rm gal}(m)$ relation, we
run a Monte Carlo with varying parameters for the $N_{\rm gal}(m)$
relation,

\beq 10^9 M_{\sun}/h \leq m_{\rm cut} \leq 10^{13} M_{\sun}/h,\ \ \ \
\ m_{\rm cut} \leq m_0 \leq m_{\rm cut} \times 10^4,\ \ \ \ \ -1 \leq
a_1 \leq 4, \ \ \ \ \ 0 \leq a_2 \leq 1.5, \eeq

\noindent and use the bias parameters from Eq.(\ref{beff}) in
Eq.(\ref{S3S4g}) to decide whether a given model is accepted. We
considered both PS and ST mass functions. For the mass, we use that at
$R=20$ Mpc/h, $\sigma^2=0.145$, $S_3=2.9$, and $S_4=14.6$. Note that
the inferred deprojected variance of the APM corresponds to a median
redshift $\bar{z}=0.15$, so we extrapolate the predictions from
$\Lambda$CDM $\sigma_8=0.90$ to this redshift.  

If we impose no further constraint on the high-mass slope $a_2$, we
find that the maximum skewness (at $R=20$Mpc/h) is $S_3=3$ for the ST
mass function (with $a_2=1$) and $S_3=3.1$ for the PS mass function
(with $a_2=1.1$). However, such a high value for $a_2$ means that
1-halo terms are not suppressed with respect to the mass, and thus the
variance and skewness at scales $R \approx 1-5$Mpc/h are much larger
than observed. If we restrict $a_2 \la 0.9$, we find that the maximum
skewness for the PS mass function becomes $S_3=2.4$, uncomfortably
small for APM galaxies (but see below for discussion of APM
deprojection issues). For the ST mass function we find that values as
high as $S_3=2.64$ are possible (with $S_4=12.6$), this requires in
addition that $a_1 \approx a_2=0.9$ and $m_{\rm cut} \la 10^{10}
M_{\sun}/h$, so galaxy formation is relatively efficient in small-mass
halos.  This values are insensitive to $m_0$, as $a_1 \approx a_2$ and
the large-scale clustering is insensitive to the distribution of
galaxies within halos. Essentially, in this model galaxies trace as
much as possible the dark matter. However, as shown in
Figs.~(\ref{fig_var_APM}-\ref{fig_Sp_APM}) in solid lines ($m_{\rm
cut}=8\times 10^9 M_{\sun}/h$, $m_0=6 \times 10^{10} M_{\sun}/h$,
$a_1=1.2$, $a_2=0.9$), the small-scale variance and skewness are
overestimated in this model. So, further suppression of galaxy
formation in high-mass halos is required, $a_2<0.9$. Similarly, the
galaxy model of Eq.(\ref{SD}) (dot-dashed lines in
Figs.~(\ref{fig_var_APM}-\ref{fig_Sp_APM})) suffers from the same
problem for the skewness.

Before we turn to constraints derived from small-scale clustering,
we should note that these depend on at least two additional assumptions 
(rather than just the $N_{\rm gal}(m)$ relation). 
First, we are assuming that galaxies trace the dark matter profile. 
Fig.~2 in Diaferio et al. (1999) shows that this cannot be true for 
both red and blue galaxies.  The blue semianalytic galaxies, which 
should be more like the ones in the APM survey, are preferentially 
located in the outer regions of their parent halos.  In the semianalytic 
models, this happens because, in the time it takes for a galaxy's orbit 
to decay (by dynamical friction) from the edge of its parent halo to 
the centre, its stars age, so its colour changes from blue to red. 
Blue galaxies on approximately radial orbits which might pass close 
to the halo centre spend most of their time far from the centre 
anyway.  As a result, most galaxies near the halo centre are red, 
and those further out are blue.  To mimic this effect, we have studied 
what happens to our model predictions if we decrease the concentration 
parameter by a factor of two; as explained above, this decreases the 
variance but it does not change the ratio of moments such as the 
skewness appreciably, so our results seem robust to this effect. 

Second, recall that we are assuming that halo-halo correlations are
well described by leading order PT. In the case of the clustering of
dark matter, this was well justified because at scales where exclusion
effects become important (invalidating the extrapolation from PT),
1-halo terms dominate over halo-halo correlation contributions.  For
galaxies this may not be true anymore, particularly since many small
mass halos host at most one galaxy, so the 1-halo term from these
halos is suppressed. For example, we find that at 3~Mpc/h, the
contributions of 1-halo and 2-halo terms to the variance and the third
moment of the galaxy counts are comparable, whereas this scale is
about 5~Mpc/h for the dark matter.  If one assumes that exclusion
effects suppress the contribution of 2-halo terms (see e.g. Fig.~5 in
Mo \& White 1996, or the analytic treatment of exclusion effects in
Sheth \& Lemson 1999), one might conclude that the skewness is {\em
higher} than what we get by including 2-halo contributions using PT,
since these affect more the square of the variance than the third
moment.  On the other hand, Mo, Jing \& White (1997) show that, on
scales smaller than the non-linear scale, our formulae for the
3-halo and 4-halo terms significantly overestimate the skewness and
kurtosis of the haloes in their simulations (see the bottom right
panels of their Figs.~3 and~4).  Therefore, although these exclusion
effects may conspire to approximately cancel out in the end, we should
bear in mind that non-trivial behavior from exclusion effects can
invalidate our conclusions from clustering statistics at small scales.

Modulo these caveats, if we impose the additional constraint that at
small scales, e.g. $R=1$Mpc/h, $(S_3)_{\rm gal} \la 6$, we find models
with parameters

\beqa
\label{param1}
m_{\rm cut}&=&4\times 10^{9} M_{\sun}/h,\ \ \ \ \ 
m_0=8\times 10^{11} M_{\sun}/h,\ \ \ \ \ 
a_1=1,\ \ \ \ \ 
a_2=0.8, \\
\label{param2}
m_{\rm cut}&=&2.5\times 10^{10} M_{\sun}/h,\ \ \ \ \ 
m_0=10^{12} M_{\sun}/h,\ \ \ \ \ 
a_1=0.8,\ \ \ \ \ 
a_2=0.8, \\
\label{param3}
m_{\rm cut}&=&6\times 10^{10} M_{\sun}/h,\ \ \ \ \ 
m_0=1.2 \times 10^{12} M_{\sun}/h,\ \ \ \ \ 
a_1=0.6,\ \ \ \ \ 
a_2=0.8,
\eeqa

\noindent the first of which is shown in
Figs.~\ref{fig_var_APM}-\ref{fig_Sp_APM} as dotted lines (the others
have very similar behavior). However, all these models predict a small
scale variance which is too large; in fact in Fig.~\ref{fig_var_APM}
we have used a concentration parameter a factor of two smaller than
Eq.(\ref{conc}) to decrease the small-scale variance .  We were unable
to find a model which matched both the variance and skewness of the
APM survey at all scales.  Therefore, we conclude that the skewness
provides a stringent constraint on models of galaxy formation. In
particular, the relatively large skewness at large scales and
relatively small skewness at small scales provide opposite
requirements on the number of galaxies per halo for massive halos. The
``large'' value of the skewness at large scales requires that galaxies
trace mass; however the small value of the skewness at small scales
requires that galaxy formation be suppressed in massive halos.

In deprojecting from angular to three-dimensional space, Gazta\~naga
(1994,1995) assumed the validity of the hierarchical ansatz for the
three- and higher-order correlation functions. At large scales,
however, PT predicts that the hierarchy of correlation functions is
not a simple hierarchical model with constant amplitudes, but rather
the amplitudes depend strongly on the shape of the configuration (i.e.
the bispectrum depends on the triangle configuration). This can affect
the deprojection (Bernardeau 1995; Gazta\~naga \& Bernardeau 1998); in
particular it can lower the three-dimensional skewness and
higher-order moments deduced from the angular data, perhaps by as much
as 20\% at $R \ga 10$Mpc/h (Gazta\~naga, private communication).  At
large scales, at least, this would improve the agreement with
halo-model predictions.

\section{Conclusions} 

We used the formalism of Scherrer \& Bertschinger (1991) to construct
an analytic model of the dark matter and of galaxies. For the dark
matter, we find that the predictions are in good agreement with
numerical simulations, except for the configuration dependence of the
bispectrum at small scales. This (numerically small) disagreement can
be traced to the assumption that halo profiles are spherically
symmetric; in reality the halos generically found in CDM simulations
are triaxial.  In general, halo models provide accuracy no better than
$20\%$ when compared to simulations.  However, as N-body results
converge on the values of the ingredients of halo models (profiles,
mass functions, etc.), their predictions will improve.

We showed how the finite volume of the simulation box can
significantly affect the results of higher-order statistics, due to
the fact that small boxes have a deficit of massive halos.  In
particular, we showed that, by making suitable cuts in halo mass
function, we were able to reproduce the observed behavior of the
bispectrum in small-box (100 Mpc/h) simulations. At small scales, halo
models predict a significant departure from the hierarchical scaling,
unless the low-mass dependence of the concentration parameter, the
mass function, and the small-scale slope of the halo profile are
different from the currently accepted values. Unfortunately, the
limited resolution of our simulations cannot test these
predictions. We caution that Ma \& Fry's (2000) conclusion about the
breakdown of the hierarchical ansatz in numerical simulations is
premature as their results are likely to suffer from finite volume
effects and inadequate resolution.

If galaxies in individual dark matter halos trace the dark matter
profiles, galaxy clustering is completely determined within halo
models by specifying the moments of galaxy counts as a function of
halo mass.  In general, suppression of galaxy formation in large-mass
halos leads to a power-law like behavior for the galaxy power spectrum
and higher-order moments which are smaller than for the dark matter.
This is similar to what is observed in galaxy surveys.  However, a
quantitative comparison with counts-in-cells statistics in the APM
survey puts stringent constraints on the galaxy counts as a function
of halo mass. At large scales, these require that galaxies trace the
mass as closely as possible, implying that galaxy formation is
relatively efficient even in small-mass halos. In addition, the
small-scale behavior of the skewness requires a high-mass slope for
the $N_{\rm gal}(m)$ relation of about $a_2=0.8$, although we found no
model which simultaneously fits the small scale value of the second
moment. The parameters of our ``best fit models'' are given in
Eqs.~(\ref{param1}-\ref{param3}). These constraints are independent of
those derived from the luminosity function and Tully-Fisher relation
which are sensitive to the overall amplitude of the galaxy counts as a
function of halo mass.

Clearly, halo models provide a useful framework within which to
address many interesting questions about galaxy clustering, and also
related topics such as galaxy-galaxy and quasar-galaxy lensing. Our
treatment should be of particular interest for the interpretation of
clustering in future galaxy surveys, such as SDSS and 2dF. In these
cases, however, redshift distortions of clustering must also be taken
into account. We plan to address this issue and others relevant to
upcoming galaxy surveys in the near future.

\acknowledgments

\begin{figure}[t!]
\centering
\centerline{\epsfxsize=18. truecm \epsfysize=15. truecm 
\epsfbox{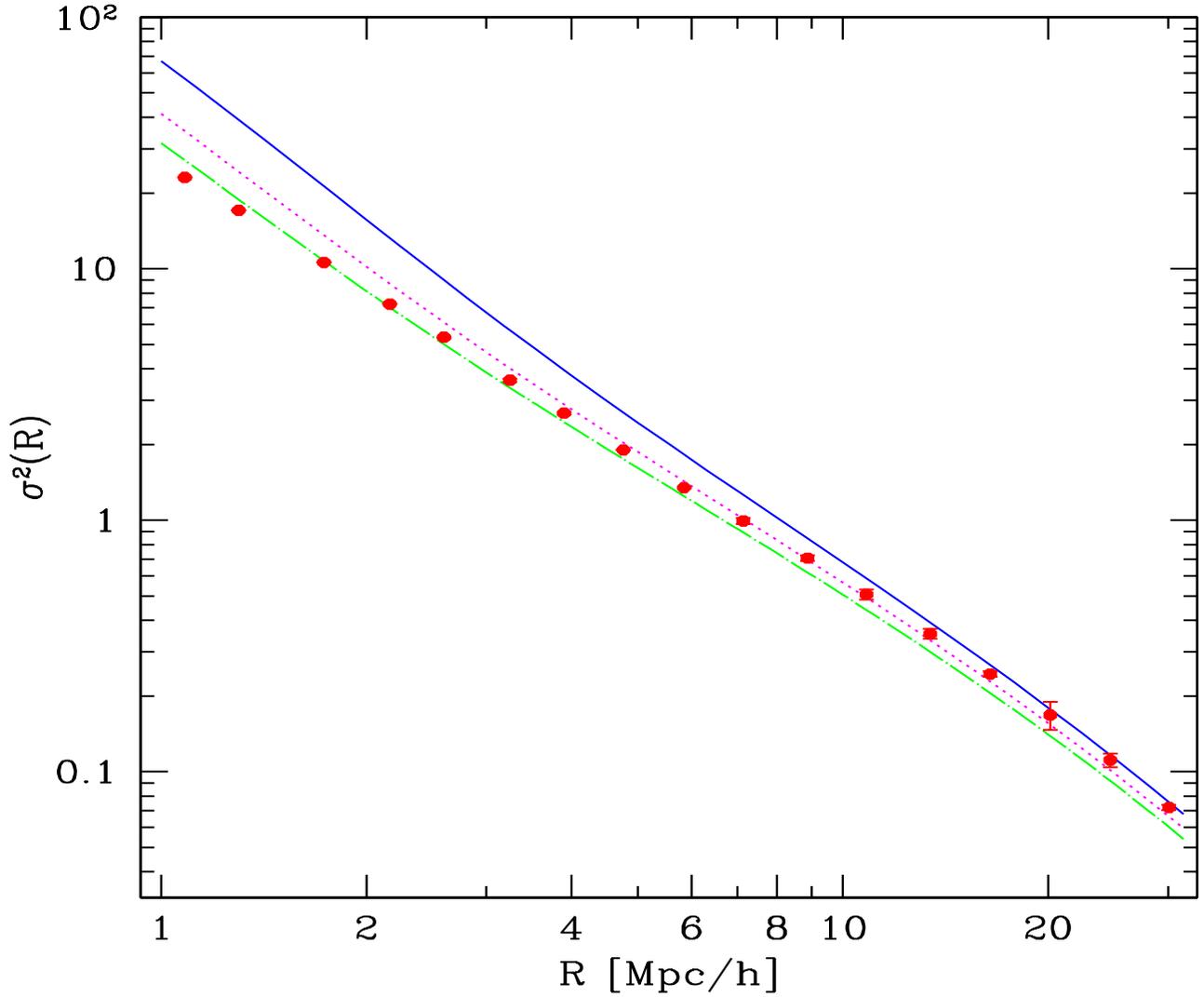}}
\caption{The variance of APM galaxies as a function of smoothing scale
$R$ (symbols) compared to the predictions for galaxies as in
Eq.(\protect\ref{SD}) with $m_{\rm cut}=10^{11} M_{\sun}$/h
(dot-dashed) and galaxies from Eq.(\protect\ref{Ngcons}) (solid) with
$a_1=1.2$, $a_2=0.9$, $m_{\rm cut}=0.8\times 10^{10} M_{\sun}/h$ and
$m_0=6 \times 10^{10} M_{\sun}/h$. Dotted lines show the predictions
for using Eq. (\protect\ref{Ngcons}) with the parameter values in
Eq. (\protect\ref{param1}).}
\label{fig_var_APM}
\end{figure}

\begin{figure}[t!]
\centering
\centerline{\epsfxsize=18. truecm \epsfysize=15. truecm 
\epsfbox{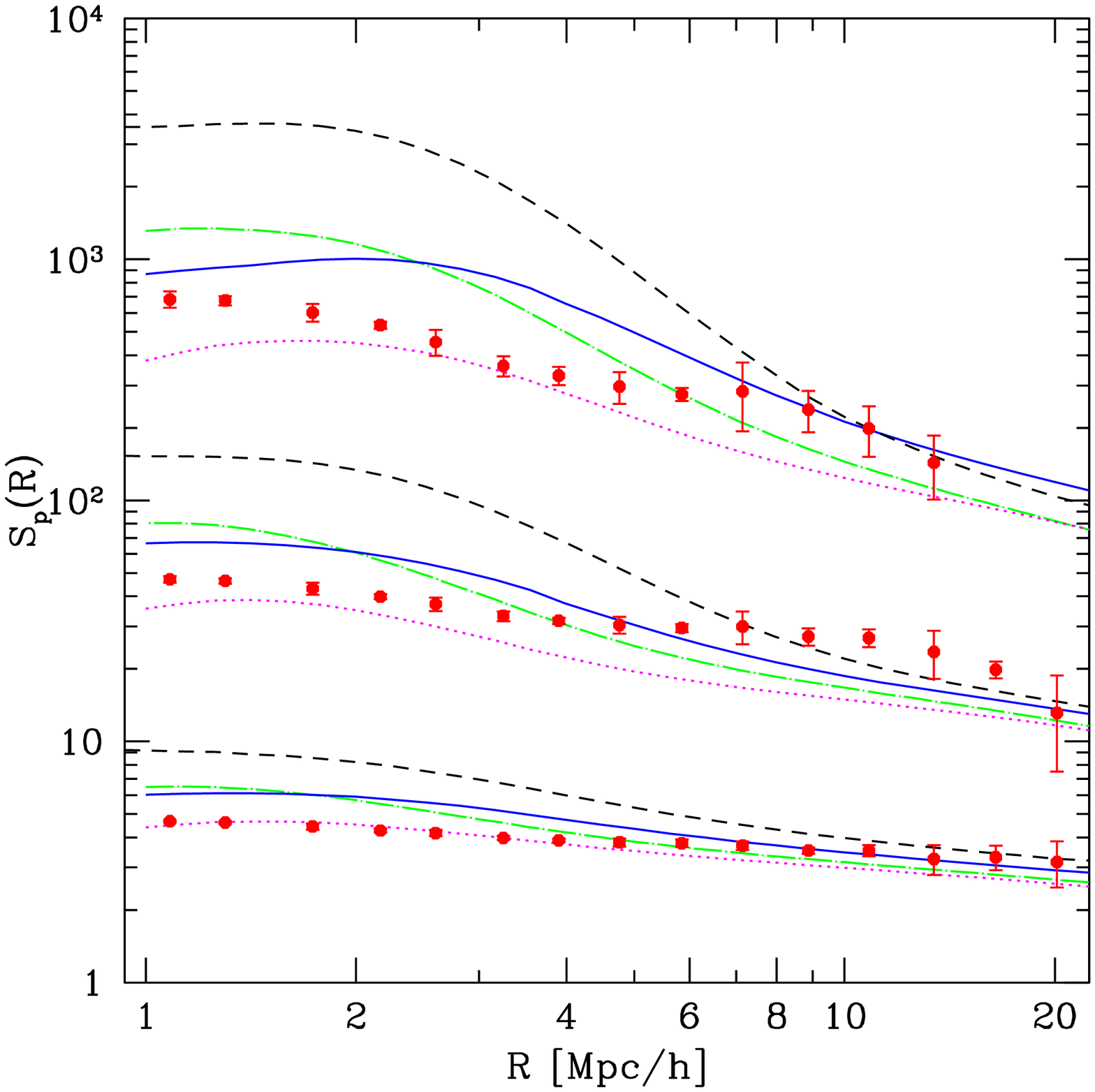}}
\caption{Same as previous figure for the $S_p$ parameters.  In
addition, the dashed lines show the predictions for the mass.}
\label{fig_Sp_APM}
\end{figure}

We thank Antonaldo Diaferio for helpful discussions about the
semianalytic models. We thank Enrique Gazta\~naga for making available
the data shown in Figs.~\ref{fig_var_APM}-\ref{fig_Sp_APM}, for
discussions on higher-order statistics from the APM survey, and useful
comments on an earlier version of this paper. We also thank Uros
Seljak for discussions about galaxy clustering.  The N-body
simulations generated for this work were produced using the Hydra
$N$-body code (Couchman, Thomas, \& Pearce 1995). We thank R.~Thacker
with help regarding the use of Hydra.  Special thanks are due to the
Halo Pub for providing much needed diversions.  This collaboration was
started in 1998 during the German-American Young Scholars Institute in
Astroparticle Physics, at the Aspen Center for Physics and MPI. We
thank Simon White for discussions and encouragement. L.H. and
R. Scoccimarro thank Fermilab and R. Sheth thanks IAS for hospitality,
where parts of this work were done. L.H. is supported by the NASA
grant NAG5-7047, the NSF grant PHY-9513835 and the Taplin
Fellowship.B.J. is supported by an LTSA grant from
NASA. R. Scoccimarro is supported by endowment funds from the
Institute for Advanced Study. R. Sheth is supported by the DOE and
NASA grant NAG 5-7092 at Fermilab.

\clearpage

\clearpage

\begin{references}

Barnes, J., \& Efstathiou, G. 1987, \apj, 319, 575

Benson, A.J., Cole, S., Frenk, C.S., Baugh, C.M., \& Lacey, C.G. 2000,
\mnras, 311, 793

Bernardeau F. 1994, \apj, 433, 1

Bernardeau F. 1995, A\&A, 301, 309

Bullock, J.S., Kolatt, T.S., Sigad, Y., Somerville, R.S., Kravtsov,
A.V., Klypin, A.A., Primack, J.R., Dekel, A. 1999, {\sf
astro-ph/9908159} 

Catelan, P., Lucchin, F., Matarrese, S., \& Porciani, C. 1998, \mnras,
297, 692 

Colombi, S., Bouchet, F. R., \& Hernquist, L. 1996, \apj, 465, 14

Cooray, A., \& Hu, W. 2000, {\sf astro-ph/0004151}

Couchman H. M. P., Thomas P. A., Pearce F. R. 1995, 
\apj, 452, 797

Diaferio, A., Kauffmann, G., Colberg, J.M., \& White, S.D.M. 1999,
\mnras, 307, 537 

Durrer, R., Juszkiewicz, R., Kunz, M., \& Uzan, J-P. 2000, {\sf
astro-ph/0005087} 

Frenk, C.S., White, S.D.M., Davis, M., \& Efstathiou, G. 1988, \apj,
327, 507 

Frieman, J., \& Gazta\~naga, E. 1994, \apj, 425, 392

Frieman, J., \& Gazta\~naga, E. 1999, \apj, 521, L83

Fry, J. N. 1984, ApJ, 279, 499 

Fry J. N., Gazta\~naga E. 1993, \apj, 413, 447

Fry, J. N. 1994, \prl, 73, 215 

Fry J. N., Scherrer, R. 1994, \apj, 429, 36

Gazta\~naga, E. 1994, \mnras, 268, 913

Gazta\~naga, E. 1995, \apj, 454, 561
 
Gazta\~naga, E., M\"ah\"onen, P. 1996, \apj, 462, L1
 
Gazta\~naga, E., Bernardeau F. 1998, A\&A, 331, 829
 
Gazta\~naga, E., Fosalba, P. 1998, \mnras, 301, 524 
 
Hamilton, A. J. S., Kumar, P., Lu, E., \& Matthews, A. 1991, \apj,
374, L1  

Hui, L., \& Gazta\~naga, E., 1999, \apj, 519, 622

Jain, B. 1997, \mnras, 287, 687 

Jain, B., Mo, H.J. \& White, S. D. M. 1995, \mnras, 276, L25 

Jenkins, A., Frenk, C.S., White, S.D.M., Colberg, J.C., Cole, S.,
Evrard, A.E., Yoshida, N. 2000, {\sf astro-ph/0005260}

Jing, Y. P., Mo, H. J., \& B\"orner, G. 1998, ApJ  , 494, 1

Kauffmann, G., Colberg, J.M., Diaferio, A., \& White, S.D.M. 1999,
\mnras, 303, 188 

Ma, C-P, \& Fry, J.N. 2000, {\sf astro-ph/0003343}

Ma, C-P, \& Fry, J.N. 2000b, {\sf astro-ph/0005233}

Maddox, S.J., Efstathiou, G., Sutherland, W.J., \& Loveday, L. 1990,
\mnras, 242, 43

McClelland, J., \& Silk, J. 1977, ApJ, 216, 665 

McClelland, J., \& Silk, J. 1977b, ApJ, 217, 331 

McClelland, J., \& Silk, J. 1978, ApJS, 36, 389 

Matarrese, S., Verde, L., \& Heavens, A.F. 1997, \mnras, 290, 651 

Mo, H. J., \& White, S. D. M. 1996, MNRAS, 282, 347 
   
Mo, H. J., Jing, Y. P., \& White, S. D. M. 1997, MNRAS, 284, 189 

Moore, B., Quinn, T., Governato, F., Stadel, J., \& Lake, G. 1999,
\mnras, 310, 1147

Navarro, J. F., Frenk, C. S., \& White, S. D. M. 1996, ApJ, 462, 563

Navarro, J. F., Frenk, C. S., \& White, S. D. M. 1997, \apj, 490, 493 

Neyman, J., \& Scott, E.L. 1952, \apj, 116, 144

Peacock, J. A. \& Dodds, S. J. 1996, \mnras, 280, L19 

Peacock, J. A. \& Smith, R.E. 2000, {\sf astro-ph/0005010} 

Peebles, P.J.E. 1974, A\&A, 32, 197

Peebles, P.J.E., 1980, The Large Scale
Structure of the Universe. Princeton University Press, Princeton

Press, W.H., \& Schechter, P. 1974, \apj, 187, 425 

Scherrer, R. J., \& Bertschinger, E. 1991, ApJ, 381, 349 

Scoccimarro, R., Colombi, S., Fry, J.N., Frieman, J.A., Hivon, E., \&
Melott, 1998, \apj, 496, 586

Scoccimarro, R., \& Frieman, J.A. 1999, \apj, 520, 35

Scoccimarro, R. 2000, {\sf astro-ph/0004086} 

Scoccimarro, R., Feldman, H.A., Fry, J.N., \& Frieman, J.A. 2000, {\sf
astro-ph/0004087}

Seljak, U. 2000, {\sf astro-ph/0001493} 

Sheth, R. K., \& Jain, B. 1997, \mnras, 285, 231 

Sheth, R. K. 1996, MNRAS, 281, 1124 

Sheth, R. K., \& Lemson, G. 1999, \mnras, 304, 767 

Sheth, R. K., \& Tormen, B. 1999, \mnras, 308, 119 

Sheth, R. K., \& Diaferio, A. 2000, in preparation

Somerville R. S., Primack J. R., 1999, MNRAS, 310,1087

Szapudi I.,  Colombi S. 1996,  \apj, 470, 131

Szapudi I.,  Colombi S.,  Jenkins A., Colberg J., 1999, submitted to
MNRAS, {\sf astro-ph/9912238}.
 
White, S. D. M., \& Frenk, C. S., 1991, \apj, 379, 52

White, S. D. M., \& Rees, M. J., 1978, \mnras, 183, 341

Zurek, W.H., Quinn, P.J., \& Salmon, J.K. 1988, \apj, 330, 519
\end{references}
\end{document}